\documentclass{aa}  

\usepackage{graphicx}
\usepackage{natbib}
\usepackage{siunitx}
\usepackage{multirow}
\usepackage[varg]{txfonts}
\usepackage{amsfonts, amsmath, amssymb}
\bibpunct{(}{)}{;}{a}{}{,} 
\begin{document}

   \title{Multiband embeddings of light curves}
   \author{I. Becker
          \inst{1, 2, 3}
          \and
          P. Protopapas\inst{1}
          \and
          M. Catelan \inst{2, 4, 5}
          \and
          K. Pichara \inst{3}
          }
    \institute{John A. Paulson School of Engineering and Applied Sciences, Harvard University, Cambridge, MA, 02138
    \email{iebecker@fas.harvard.edu}
    \and
    Millennium Institute of Astrophysics, Nuncio Monse\~{n}or Sotero Sanz 100, Of. 104, Providencia, Santiago, Chile
    \and
    Computer Science Department, School of Engineering, Pontificia Universidad Cat\'olica de Chile, Av. Vicu\~na Mackenna 4860, 7820436 Macul, Santiago, Chile
    \and
    Instituto de Astrofísica, Facultad de Física, Pontificia Universidad Católica de Chile, Av. Vicu\~na Mackenna 4860, 7820436 Macul, Santiago, Chile
    \and 
    Centro de Astroingeniería, Pontificia Universidad Católica de Chile, Av. Vicu\~na Mackenna 4860, 7820436 Macul, Santiago, Chile
    }

   \date{Received July 17, 2023; accepted January 9, 2025}

 
  \abstract{
    In this work, we propose a novel ensemble of recurrent neural networks (RNNs) that considers the multiband and non-uniform cadence without having to compute complex features. Our proposed model consists of an ensemble of RNNs, which do not require the entire light curve to perform inference, making the inference process simpler. The ensemble is able to adapt to varying numbers of bands, tested on three real light curve datasets, namely {\em Gaia}, Pan-STARRS1, and ZTF, to demonstrate its potential for generalization. We also show the capabilities of deep learning to perform not only classification, but also regression of physical parameters such as effective temperature and radius. Our ensemble model demonstrates superior performance in scenarios with fewer observations, thus providing potential for early classification of sources from facilities such as Vera C. Rubin Observatory's LSST. The results underline the model's effectiveness and flexibility, making it a promising tool for future astronomical surveys. Our research has shown that a multitask learning approach can enrich the embeddings obtained by the models, making them instrumental to solve additional tasks, such as determining the orbital parameters of binary systems or estimating parameters for object types beyond periodic ones.
   }

   \maketitle
%

\section{Introduction}
Variable objects have a unique place in astronomy, as they provide additional information which is absent in the non-variable counterparts. They allow the measurement of distances, the study of physical processes that occur inside the objects, and the study of stellar populations \citep[for a review, see][]{Catelan2015}.
Classifying these variable objects into distinct types has posed a recurring challenge in astronomy. In the past decades, the landscape of light curve acquisition has been transformed with the advent of numerous observatories \citep[for a review, see][]{2023MmSAI..94d..56C}, altering the methodology employed. Traditionally, light curves were obtained primarily in a single band to maximize temporal coverage. However, this approach led to traditional classifiers relying on hand-crafted features specific to a single band, often incurring high computational expenses and necessitating the computation of the entire light curve. Consequently, the process became computationally intensive. Various Python packages have been developed to streamline this task, such as Feature Analysis for Time Series \citep[FATS,][]{Nun2015}, offering automation capabilities.

Instead of multiple observations in one band, most modern telescopes have been observing in multiple bands, such as {\em Gaia} \citep{vallenari2022gaia}, the Panoramic Survey Telescope and Rapid Response System \citep[Pan-STARRS,][]{2016arXiv161205560C}, the Zwicky Transient Facility \citep[ZTF,][]{Bellm_2019}, the Sloan Digital Sky Survey's \citep[SDSS, ][]{2023ApJS..267...44A} Stripe 82 \citep{2007AJ....134..973I}, the Asteroid Terrestrial-impact Last Alert System \citep[ATLAS,][]{Tonry_2018}, the SkyMapper survey \citep{onken2019skymapper}, among others. Modern single-band telescopes are still operating, such as the Gravitational-wave Optical Transient Observer \citep[GOTO, ][]{2018SPIE10704E..0CD} or the All-Sky Automated Survey for SuperNovae \citep[ASAS-SN, ][]{2014ApJ...788...48S}, as the generated light curves are of great scientific importance.

A revolution will begin in astronomy when the Vera C. Rubin Observatory Legacy Survey of Space and Time \citep[LSST,][]{2019ApJ...873..111I} begins science operations. It will scan the sky every three days, observing using one of six filters, and produce 20 TB of data per night. The relatively high cadence of observations will enable the differentiation of various variable phenomena, including transients, periodic sources, and moving sources. 
Any time a variable event is detected with respect to a reference image by the Rubin/LSST image pipeline, an alert will be distributed to the Alert Brokers \citep{bellm2019plans}. It is estimated that $\sim 10$ million alerts will be generated per night. 

Brokers are systems that ingest, process, and distribute alerts to the scientific community. The seven brokers are Automatic Learning for the Rapid Classification of Events \citep[ALeRCE,][]{2021AJ....161..242F}), AMPEL \citep{nordin2019transient}, ANTARES \citep{Matheson_2021}, BABAMUL, Fink \citep{10.1093/mnras/staa3602}, Lasair \citep{Smith_2019} and Pitt-Google\footnote{\url{https://pitt-broker.readthedocs.io/en/latest/}}.

One of the most important roles of brokers is to classify the alerts into different classes. One of the main challenges is the multiband nature of the light curves since an observation is only done in one filter at a time. There is no straightforward way to combine the information into a unified representation informative enough to classify with a few observations. Additionally, any solution has to be fast to process the alerts in real-time and needs to adapt to different observing strategies. This is an open problem that has yet to be solved conclusively. 
The training and deployment of a classifier is even more challenging at the start of the survey, when only a small number of observations per object are available, and training data is scarce.

In recent years, deep learning approaches have emerged as an alternative to feature-based methods. These approaches automatically extract their own representations, eliminating the reliance on predefined features \citep[for a review, see][]{2023RSOS...1021454S}. By avoiding the need for handcrafted features, deep learning methods tend to provide faster and less computationally expensive solutions. However, to date, these methods have not surpassed feature-based Random Forest \citep[RF,][]{Breiman2001} approaches \citep[e.g.][]{2021AJ....161..242F}. This highlights the fact that expert knowledge contained in the features is critical in the development of any automated classifier. As such, even deep learning methods can benefit from domain-specific design.

Convolutional neural networks (CNNs) have been widely used to analyze multiband time series data. However, one of the primary challenges in this domain is to construct appropriate input representations for these networks. \citet{pasquet2019pelican} tackled this challenge by transforming supernova (SN) multiband time series into a 2D representation. Their approach organized the magnitudes in different bands as rows and represented the corresponding time information as columns. To extract meaningful features from unlabeled data, they employed an encoder-decoder architecture. This representation was then utilized to classify simulated Rubin/LSST light curves. Similarly, \citet{brunel2019cnn} developed a one-dimensional CNN classification model based on the inception module \citep{szegedy2015going}. Their study focused on simulated data captured at a daily cadence across four bands. They also adopted the same matrix representation, filling in missing values with zeros.

In \citet{muthukrishna2021real}, a CNN was developed to detect anomalies in SN light curves. Training was carried out using simulated data, and testing was performed on real ZTF light curves. Linear interpolation addressed missing observations, followed by binning to achieve a cadence of three days.

The interpolation or binning approach to handle missing observations can pose challenges due to limited data points or large gaps between observations that exceed the characteristic time of the object. This effect becomes more pronounced with an increasing number of bands, as it requires more interpolations. Moreover, the number of observations per band is typically non-uniform, as they adhere to broader requirements set by the science objectives \citep{Bianco2022} and the meteorological conditions.

Recurrent neural networks (RNNs) can naturally handle sequences, utilizing learned weights and the last hidden states to update embeddings and perform classification. However, RNNs tend to have slower training compared to their CNN counterparts, as they require sequential training. This sequential nature becomes advantageous during inference, as RNNs only need the last hidden state and the new observations, rather than the entire light curve, to make predictions. Despite some benefits, the Balanced Random Forest \citep[BRF,][]{chen2004using} is used in state-of-the-art classification scenarios, such as the ALeRCE broker.

\citet[][]{Becker2020} introduced a model based on Gated Recurrent Units \citep[GRU; ][]{cho2014learning} for analyzing single-band light curves. Building on this work, \citet{donoso2021effect} enhanced the methodology employing a combination of Long Short-term Memory \citep[LSTM; ][]{Hochreiter1997} and Phased-LSTM \citep{2016arXiv161009513N}, resulting in promising results across six real datasets. Both approaches demonstrated the ability of RNNs to work without interpolation.
In another study by \citet{2017ApJ...837L..28C}, a deep bidirectional LSTM network was trained on simulated multiband light curves of supernovae. To address sparsity, the authors binned the light curve into 1-hour windows to ensure simultaneous observations. To impute missing values, they sampled random values between the observed points.

In the work of \citet{moller2020supernnova}, a Bayesian RNN was employed to classify simulated SN light curves and obtain uncertainty in the predictions. The observations were grouped into 8-hour windows to handle the irregular cadence, and the multiband information was incorporated by using one-hot-encoded vectors and included the time difference between observations.

In a ground-breaking work, \citet{vaswani2017attention} introduced the idea of self-attention and applied it to train an encoder-decoder called Transformer, removing the need for RNNs. The main benefit of this architecture is that it leverages the entire sequence in parallel, enabling the model to extract informative representations for each element in the sequence\footnote{
Language Models have been developed using this idea, obtaining state-of-the-art results trained on a massive amount of unlabeled data \citep{devlin2018bert,  brown2020language, touvron2023llama}}.

The same concept has been employed in extracting light curve representations, as demonstrated by ASTROMER \citep{2023A&A...670A..54D}, which uses a pre-training strategy on unlabeled data. Recent studies, such as \citet{2023AJ....165...18P}, have utilized self-attention architectures for the multiband classification of supernovae. The models were tested on both synthetic and real ZTF data. Their approach for constructing the input involved binning the observations into 12-hour windows and concatenating the band information as a one-hot vector.
\citet{2021arXiv210506178A} developed a model based on self-attention to classify simulated multiband light curves of variable, transient, and stochastic objects. They employed interpolation using a Gaussian process to fill in the missing observations and utilized a global average pooling layer to explain the model's output.

Supervised training can also be extended to regression tasks. Various relations involving light curve information and astrophysical quantities have been developed throughout history. For example, the Leavitt law \citep{1912HarCi.173....1L} established a relation between Cepheids' light curve properties and their absolute magnitude. In a study by \citet{1992AJ....103.1647F}, a relation between periods and radii was obtained for $\delta$ Scuti stars. Additionally, \citet{1995AJ....110.2361F} fitted a period-gravity relation for radially pulsating stars. Fourier coefficients have also been utilized to establish relations between metallicity ([Fe/H]) and pulsation period \citep{2007MNRAS.374.1421M}. Relationships of this type are not limited to pulsating stars; for instance, late-type eclipsing binaries also exhibit period-luminosity and period-luminosity-color relations \citep[][and references therein]{2021AJ....162...63N}. Deep learning methods have been applied to perform similar regressions in recent years \citep[e.g.][]{2022ApJS..261...33D}.

In this work, we propose an ensemble of RNNs that incorporates the non-uniform sampling of multiband light curves into its design to extract their representations. This approach preserves the inherent relation between different bands, irrespective of the survey's cadence. The extracted representations are obtained through training a classifier and a regression simultaneously when the data are available. This training methodology, also known as multi-task learning \citep[for a review, see][]{2020arXiv200909796C}, serves as a regularization mechanism while providing additional information to train the network. Consequently, it can enhance the performance of the models. We trained the models using three real light curve datasets, namely {\em Gaia}, Pan-STARRS1, and ZTF, to demonstrate their potential for generalization.

In Section \ref{section:Datasets}, we describe the real light curves used in this work. In Section \ref{section:Model}, we describe the single and multiband models used in this work. In Section \ref{section:Results}, we show the results of the multiband classification and the regression of physical parameters. Finally, in Sections \ref{section:Analysis} and \ref{section:Conclusion}, we present the analysis and main conclusions of this work.

\section{Datasets}\label{section:Datasets}

This work used multiple single-band light curves to represent each object. The dataset consisted of modified Julian Date (MJD) values sorted in ascending order, the corresponding magnitudes with their respective uncertainties, and the bands in which the observations were conducted. Any flux values that exceeded their associated uncertainties were discarded, to remove low S/N measurements and ensure accurate photometry.

As is common in many deep learning approaches, a large training set with an adequate number of examples per class and sufficient observations per band is necessary. To fulfill these requirements, this study utilized real photometry data from {\em Gaia}, Pan-STARRS1, and ZTF. The following sections present comprehensive descriptions and explanations of the main pre-processing steps carried out on each dataset.

\begin{figure}
\centering
\includegraphics[width=0.49\textwidth]{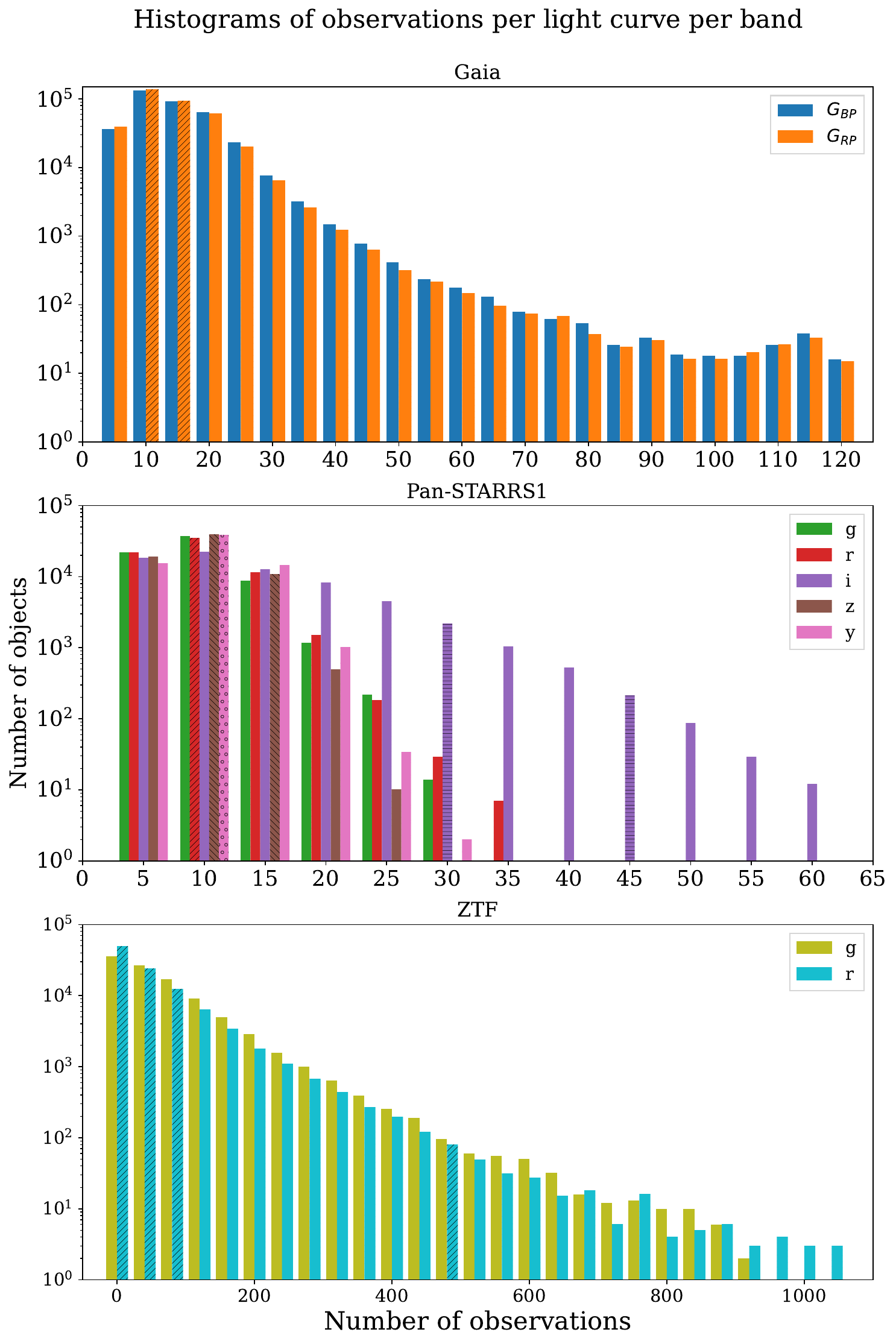}
\caption{Histograms of the number of observations on each band, per survey. Each band is identified by a color and a hatch style. {\em Gaia} filters $G_{BP}$ and $G_{RP}$ are identified with blue solid and orange hatched bars, respectively. Pan-STARRS1 bands are identified with solid green for $g$, red right diagonal hatch for $r$, purple horizontal hatch for $i$, brown left diagonal hatch for $z$, and pink dotted bar for $y$.
Both {\em Gaia} and ZTF are balanced but differ in the number of observations per light curve by one order of magnitude. Pan-STARRS1 is not balanced, showing longer sequences on the \textit{i} band. As such, not all the bands will be available for some objects. }
\label{fig:histograms_N}
\end{figure}

\begin{figure}
\centering
\includegraphics[width=0.47\textwidth]{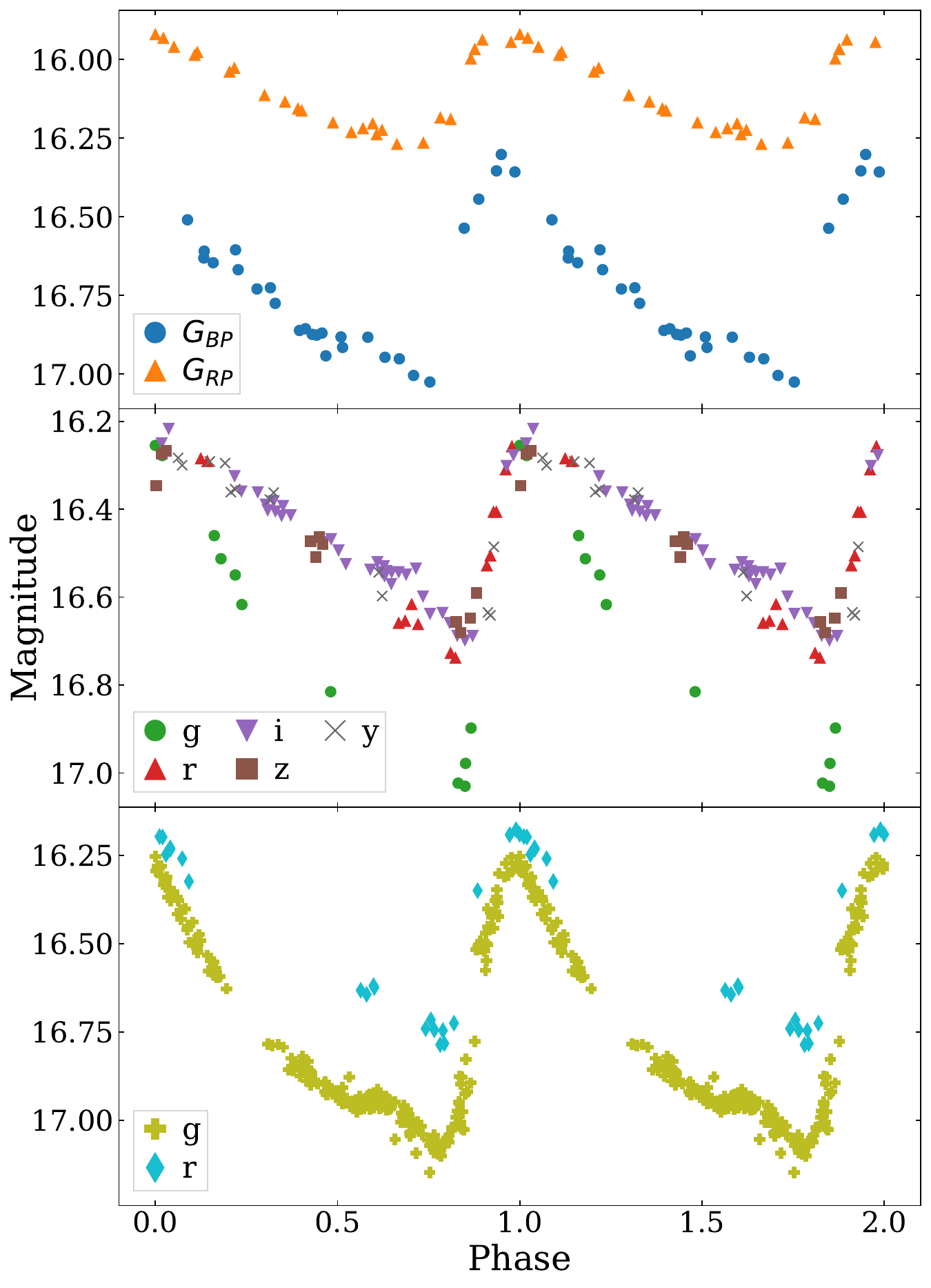}
\caption{Light curves of the same RRab star with a period of 0.658936 days. The identifiers are 839368005575354240, 169841701110866261 and ZTF17aaajjtn, for {\em Gaia} DR2, Pan-STARRS1 DR2 and ZTF DR10, respectively. 
For {\em Gaia}, the observation in the $G_{BP}$ filter is shown in blue circles and orange triangles for $G_{RP}$. For Pan-STARRS1 on the second panel, the filters \textit{$g_{P1},r_{P1},i_{P1},z_{P1},y_{P1}$} are identified by green circles, red triangles, purple inverted triangles, brown squares, and gray diagonal crosses, respectively.
For ZTF, \textit{g} is shown in olive circles and \textit{r} in cyan diamonds.
}
\label{fig:lcs}
\end{figure}

Histograms of the number of observations per light curve in each band are shown in Figure \ref{fig:histograms_N}. Figure \ref{fig:lcs} illustrates a folded light curve of an RRab star, highlighting the inconsistent number of observations across the surveys. 

\subsection{{\em Gaia}}
The {\em Gaia} DR2 variable star catalog, as described by \citet{Holl2018}, consists of \num{550737} variable objects cataloged using machine learning models. The available photometry includes three bands, with two of them being independent \citep{jordi2010gaia}. The telescope's optics divide the light beam into two, allowing simultaneous measurements of red $G_{RP}$ and blue $G_{BP}$ magnitudes. For the purposes of this study, the $G$ filter's photometry is omitted since it is redundant as $G=G_{BP}+G_{RP}$. Our objective is to obtain light curves with bands sampling different sections of the electromagnetic spectrum.

To simulate an alternating cadence, we sub-sample the real light curves. If two observations from different bands are within 15 minutes of each other, one of the observations is removed to maintain the alternating cadence. We selected objects with a minimum of four observations per band. Table \ref{tab:gaia_numbers} displays the total number of examples per class.

For this dataset, the classes used were RR Lyrae fundamental-mode (RRab) and first-overtone (RRc), Mira and semi-regular variables (MIRA\_SR); $\delta$ Scuti and SX Phoenicis (DSCT\_SXPHE); classical Cepheid (CEP) and type II Cepheid (T2CEP).

We sampled a maximum of \num{40000} objects per class to construct the final training set to prevent classifier bias. This sampling strategy primarily limits the number of RRab and MIRA\_SR stars. Categories with fewer samples than the threshold were used entirely, without any sampling.

\begin{table}
\caption[]{{\em Gaia} dataset class numbers.}         
\label{tab:gaia_numbers}     
\centering                   
\begin{tabular}{lr}          
\hline\hline                 
       Class & Number \\     
\hline                                   
        RRab & \num{161225} \\
        RRc & \num{32276} \\
        MIRA\_SR &  \num{149821} \\
        DSCT\_SXPHE & \num{8830}\\
        CEP & \num{6487} \\
        T2CEP & \num{1740} \\
        Total & \num{360379}\\       
\hline                                             
\end{tabular}
\end{table}

\subsection{Pan-STARRS1}  
Pan-STARRS1 \citep{2016arXiv161205560C, Flewelling_2020, Magnier_2020} is the first telescope of the Pan-STARRS Observatory, which is an imaging and data processing facility primarily used for the $3\pi \text{sr}$ and Medium Deep Surveys. It features a 1.8-meter telescope equipped with a 1.4 gigapixel camera, which captures observations through filters $g_{P1}, r_{P1}, i_{P1}, z_{P1},$ and $y_{P1}$. To select a sample of variable objects, we conducted a cross-match with the {\em Gaia} DR2 objects \citep{marrese2019gaia}. The classes were the same as in {\em Gaia}, with the addition of RR Lyrae double-mode pulsator (RRd).

For the second data release, the photometry was obtained from the Detections table on the MAST CasJobs\footnote{\url{https://mastweb.stsci.edu/ps1casjobs/home.aspx}}. Specifically, columns containing MJD, flux, flux uncertainty, and band information were queried.

The fluxes were transformed into AB magnitudes using
\begin{equation}
m_x^{\rm AB} = -2.5 \log_{10}{\left(\frac{F_x}{3631}\right)},\label{eq:ab_mag}
\end{equation}
where $x$ represents the band, and 3631 Jy is the zero point for the AB photometric system \citep{tonry2012pan}.

Furthermore, the columns \texttt{psfQfPerfect}, \texttt{infoFlags}, \texttt{infoFlags2}, and \texttt{infoFlags3} were extracted to ensure the selection of clean photometry. The criteria for filtering observations are outlined in Appendix \ref{appendix:PS_cleaning}.

Following the pre-processing steps, we imposed a requirement of a minimum of four observations per band. The final catalog is described in Table \ref{tab:ps_numbers}. Similar to the {\em Gaia} dataset, we sampled a maximum of \num{10000} objects per class to reduce overfitting, reducing the number of RR Lyrae objects in the dataset. The threshold value differs from the {\em Gaia} dataset because of the huge class imbalance, where RR Lyrae examples are ~$91\%$ of the total number of samples. As such, a lower value had to be selected to avoid overfitting to only one object type.

\begin{table}
\caption[]{Pan-STARRS1 dataset class numbers.}         
\label{tab:ps_numbers}     
\centering                   
\begin{tabular}{lr}          
\hline\hline                 
       Class & Number \\     
\hline                                   
        RRab & \num{51328} \\
        RRc & \num{11968} \\
        RRd & \num{266} \\
        MIRA\_SR &  \num{3937} \\
        DSCT\_SXPHE & \num{1906}\\
        T2CEP & \num{189} \\
        Total & \num{69594} \\
\hline                                             
\end{tabular}
\end{table}

\subsection{ZTF}
ZTF \citep{Bellm_2019} is an optical time-domain survey that uses the Palomar 48-inch Schmidt telescope. This telescope has a field of view of $\SI{43.56}{\deg^2}$ and uses a 576 megapixel camera. This survey uses ZTF's alert photometry in the $g$, $r$, and $i$ filters. The alert light curves were obtained from ALeRCE's Database service\footnote{\url{https://alerce.science/services/database/}} up to May 2022, where only $g$ and $r$ bands are available for the alerts. The public data releases contain all three bands.

In this work, the labeled training set objects were obtained from \citet{sanchez2021alert}, as it contains a curated sample of variable stars, transients, and stochastic variables. Transients are represented by SN of types Ia, Ibc, II and super luminous supernova (SLSN). The stochastic classes are blazars, type 1 Seyfert galaxy (AGN), type 1 quasar (QSO), young stellar object (YSO) and cataclysmic variable/nova (CV/Nova). The main categories for periodic variables are long-period variable (LPV), RR Lyrae (RRL), Cepheid (CEP), eclipsing binary (E), $\delta$ Scuti (DSCT), and Periodic-Other, which are periodic classes not considered in the previous ones.
For training purposes, we sampled up to \num{10000} objects per class, affecting mainly E, QSO, and RRL type objects.

\begin{table}
\caption[]{ZTF dataset class numbers.}
\label{tab:ztf_numbers}     
\centering                   
\begin{tabular}{lr}          
\hline\hline                 
       Class & Number \\     
\hline                                   
        SNIa & \num{1026} \\
        SLSN & \num{24} \\
        SNIbc & \num{68} \\
        SNII & \num{280} \\      
        Blazar & \num{1161} \\
        QSO & \num{23267} \\        
        YSO & \num{1052} \\       
        CV/Nova & \num{821} \\        
        AGN & \num{2946} \\        
        LPV & \num{10291} \\ 
        E &  \num{28260} \\
        DSCT & \num{553} \\
        RRL & \num{29769} \\
        CEP & \num{415} \\
        Periodic-Other & \num{795}\\
        Total & \num{100728} \\
\hline                                             
\end{tabular}
\end{table}

\subsection{Physical parameters}
Physical parameters play a crucial role as they provide fundamental descriptions of objects. However, for variable stars, these parameters often undergo changes over time. Consequently, it is only realistic to study the mean values of these parameters, when large sample sizes are involved. Otherwise, a large investment in follow-up resources would be required, which is not feasible for a large sample size.

In this study, we utilized the values obtained from the Transiting Exoplanet Survey Satellite (TESS) Input Catalog \citep[TIC,][]{TIC8.1}, which is employed by the TESS mission \citep{2015JATIS...1a4003R} for target selection, and the {\em Gaia} astrophysical parameters inference system \citep[Apsis,][]{GaiaPhys}. Both catalogs are products of systematic analyses of objects. Apsis derives stellar parameters using an ensemble of machine learning models. The physical parameters of TIC sources are mainly inferred from $T_{\rm eff}$, which is obtained from a spline fit as a function of $G_{BP}-G_{RP}$ based on spectroscopic data. The radius is derived from the Stefan–Boltzmann relation, and the mass is obtained from a fitted mass versus $T_{\rm eff}$ spline function. The rest of the quantities are obtained directly. 

After performing a cross-match, if a measurement was available in both catalogs, the measurement from the TIC took precedence as it reported smaller uncertainties in its estimations. It is worth noting that, while a more extensive search might have yielded additional measurements, we intentionally maintained a smaller set of sources to avoid introducing unnecessary sources of uncertainty.

Given the limitations of photometric and spectral observations, not every object will have measurements for these parameters, let alone measurements for all of them.

This model utilized the reported central values without considering the uncertainties, which are likely underestimated for TIC. In its current form, the regression was only used to enhance the information contained in the embeddings and improve the classification performance.

\begin{table}
  \caption[]{Physical parameters measured on each dataset.}
     \label{tab:physical_parameters}    
\centering                   
\begin{tabular}{lrrr}          
\hline\hline                 
        Parameter      &  \text{{\em Gaia}} & \text{Pan-STARRS1}& \text{ZTF}\\     
\hline                                   
    $T_{\rm eff}$ & \num{162650} &\num{48932}&\num{65619}\\
        Radius & \num{140027} &\num{40956}&\num{55686}\\
\hline                                             
\end{tabular}
\end{table}

Table \ref{tab:physical_parameters} provides an overview of each survey's number of objects with measurements of $T_{\rm eff}$ and radius. To avoid numerical artifacts, we kept values of $T_{\rm eff}$ between 3400 K and 8000 K, and radii between 10 and 200 $\text{R}_\odot$.

\section{Multiband RNN classifier}\label{section:Model}
In this work, we built on the ideas proposed by \citet{Becker2020} to create a novel model inspired by the Local-Global Hybrid Memory Architecture \citep{2016arXiv160907222L}. This model incorporates modular single-band interconnect components to facilitate information flow, thereby increasing flexibility and adaptability.

A single multiband model's limitation lies in its inflexibility to include new information post-training. To overcome this, we developed a model that uses the neural network's capacity to generate its own representations and harnesses its modular nature to extend the architecture, integrating new filters into the data. Consequently, the model can adapt to different surveys, accommodate additional bands, and even assimilate data from various telescopes.

We constructed our proposed model from interlinked single-band models based on LSTMs. Each of these models uniquely learns the behavior of the light curves in its respective band. These single-band models feed into another LSTM, which then develops a unified multiband vector representation. We used this consolidated representation for subsequent tasks, such as variable star classification and the regression of stellar physical parameters.

This section provides a detailed description of the pre-processing stage, the creation of single-band inputs, the architecture of the model that learns the unified representation, and the training strategy employed to train the model effectively.

\subsection{Pre-processing}\label{section:preprocessing}

Our encoding of single-band light curves followed a similar method to \citet{Becker2020}, with some additional steps. We used an integer $b$ to identify the band and to store the uncertainties $\vec{u}^b$ related to the measurements. We assigned each time measurement in the multiband light curve an integer denoting the observation order, which we stored in the vector $\vec{o}^b$.

For each band, we calculated the differences in time and magnitude between the current and previous observations, excluding the first observation. We then grouped these differences using a sliding window of size $w$ and stride $s$. In this work, we set $w=2$ to allow predictions from the third observation and $s=1$ to predict after each subsequent observation. We maintained these values across all datasets. While this representation omitted the first observation per band, the difference became negligible as we accumulated sufficient observations.

The resulting single-band input $\textbf{X}^b$ is,
    \begin{equation}
    \label{eq:matrix_representation}
    \textbf{X}^b = \left (
    \begin{bmatrix}
    \Delta t_2  &\Delta t_3 & \Delta m_2& \Delta m_3\\
    \Delta t_3  &\Delta t_4 & \Delta m_3& \Delta m_4\\
    \vdots & & &\vdots \\
    \Delta t_{N\text{-}1}  &\Delta t_N & \Delta m_{N\text{-}1}& \Delta m_N\\
    \end{bmatrix}\;,\;
    \begin{bmatrix}
    u_3 \\
    u_4 \\
    \vdots \\
    u_N
    \end{bmatrix}
    \;,\;
    \begin{bmatrix}
    o_3 \\
    o_4 \\
    \vdots \\
    o_N
    \end{bmatrix}
    \right ).
  \end{equation}
We also stored non-sequential data per band, such as the number of observations and the first measurements of time and magnitude. 

\subsection{Single-band representation}\label{subsection:single-band}
This work used a multi-layer LSTM as the primary mechanism for learning representations. We considered each row of the matrix $\boldsymbol{X}^b$ as a step in the sequence and used it as input for the $b$-th single-band model.
Additionally, we applied residual connections \citep{he2016deep}, as demonstrated in \citet{wu2016google}, between the recurrent layers to promote the flow of gradients and augment the model's expressivity. We also applied Layer Normalization \citep{ba2016layer} between each recurrent layer.

The hidden state encodes the light curve information from each recurrent layer up to that specific time step. Adding recurrent layers can incur an increased computational cost and yield diminishing returns, a pattern we observed in preliminary experiments and corroborated by previous works \citep{zhang2016architectural}. To merge the hidden states without substantially increasing the number of parameters, we used a linear combination of the hidden states, adhering to the method proposed by \citet{peters2018elmo}.

For each step $i$ in the sequence, we combined the $L$ hidden states using
\begin{equation}
\tilde{\vec{H}}_i = \sum_{l}^{L}{\alpha_l \cdot \vec{h}^{l}_{i} };
\label{eq:elmo_sauce}
\end{equation}
we softmax-normalized the trainable parameters $\alpha_l$. This normalization allowed the model to merge the information without concatenating the hidden states or adding more layers. In this work, we selected $L=3$ for all three datasets. We set the hidden state size to $128$ for {\em Gaia} and ZTF, and $64$ for Pan-STARRS1. The values for the hidden state sizes were defined based on preliminary experimentation and were not fine-tuned. The sizes correlated with the length of the light curves. A larger hidden state implied a larger number of parameters, which required more training data. Pan-STARRS1 light curves were smaller compared to the other two surveys.

Following the methodology in \citet{donoso2021effect}, we performed classification at each time step to encourage early sequence predictions by the model. We used the categorical cross-entropy loss function. However, we excluded the first $N_{skip}$ predictions to compensate for the model's limited early performance with sparse observations. We always set $N_{skip}$ to be less than the shortest sequence in the dataset. Specifically, we set $N_{skip}$ to 8 for {\em Gaia}, 2 for Pan-STARRS1, and 3 for ZTF.

Light curves include the uncertainty of each observation, which can be used to prioritize them. In a survey setting, fainter stars often have larger uncertainties due to lower signal-to-noise ratios than brighter stars. If not addressed, the model might give undue weight to certain observations tied to brighter objects.

To counteract this bias, we normalized the weights of the observations within each light curve to a total one. This process minimized the impact of uncertain observations on the predictions for each light curve and treated every example in the dataset equally. For each example, we computed a weighted average of the individual step losses. We chose the weight of each time step as the inverse of the uncertainty value. The weight of the $l$-th step for a single light curve is

\begin{equation}
w_l = \frac{u_l^{\text{-1}}}{\sum_{j=N_{skip}}^{N_l}{u_j^{\text{-}1}} }.\label{eq:uncertainty}
\end{equation}

\begin{figure*}
\centering
\includegraphics[width=0.9\textwidth]{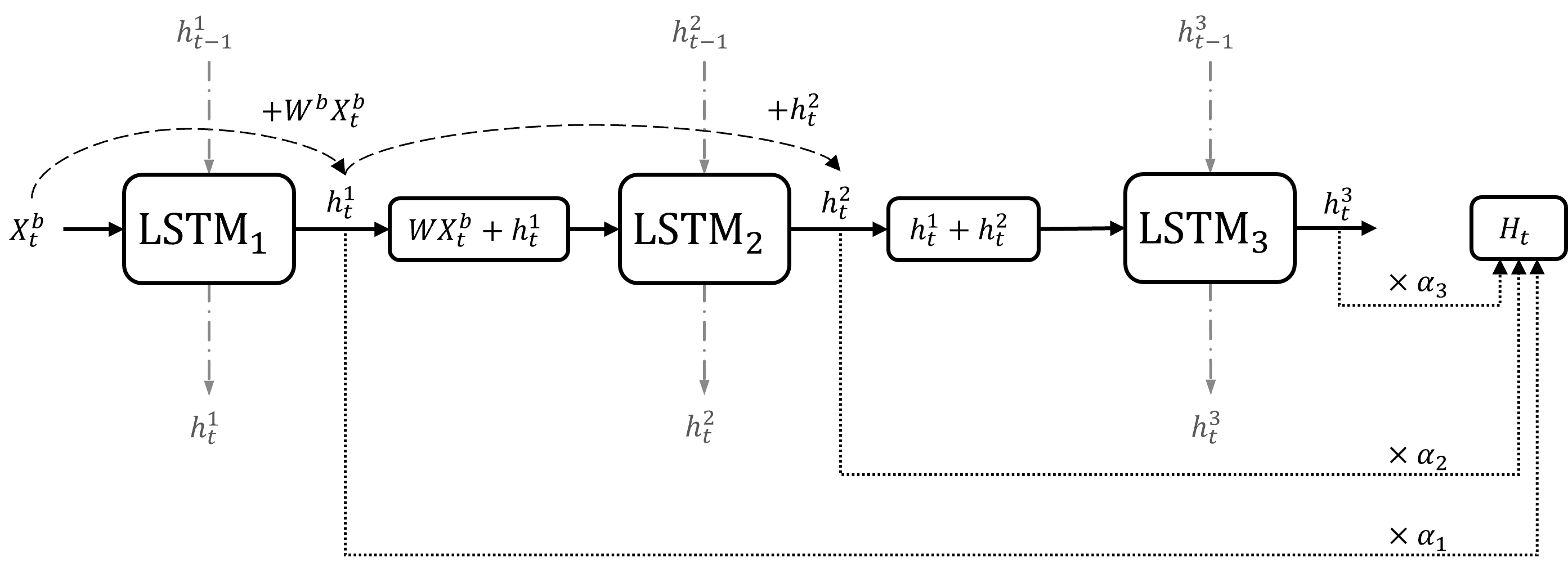}
\caption{Single-band representation of a single time step. Solid lines represent the inputs and outputs of the LSTM cells across a single time step. Dashed lines represent the residual connections between recurrent steps. The dotted lines represent the operations to construct the final representation, given in Equation \ref{eq:elmo_sauce}. The gray connections represent the inputs and outputs across different time steps. }
\label{fig:singleband_network}
\end{figure*}

\noindent In this equation, $N_l$ represents the length of the $l$-th light curve, and the subscript $j$ corresponds to the $j$-th element of the sequence. We calculated the loss of each batch as the average across the different training examples within the batch. A diagram of a single-band model is shown in Figure \ref{fig:singleband_network}.

\subsection{Multiband representation}\label{subsection:multiband}

We employed an additional LSTM to merge information from the single-band representations, whose architecture is similar to the single-band ones. The inputs to this network were the hidden states of each single-band LSTM, sorted using the order information $\vec{o}$ detailed in Sect. \ref{section:preprocessing}.

We applied a distinct linear combination, using Eq. \ref{eq:elmo_sauce}, to create a unified vector input for each band. The central model learned the weights of this transformation, enabling it to attend to the hidden states in a manner distinct from the single-band models.

Given that the single-band RNNs are trained independently, the embeddings $\tilde{\vec{H}}$ cannot be directly compared. To overcome this, we trained a three-layer Feed Forward Neural Network (FFNN) with Rectified Linear Unit (ReLU) activation to map the outputs of each single-band RNN to a new representation $\tilde{\vec{S}}$. We refer to these FFNNs as translation layers, and their weights are trained by the central model. After this step, we applied Layer Normalization.

\begin{figure*}
\centering
\includegraphics[width=0.9\textwidth]{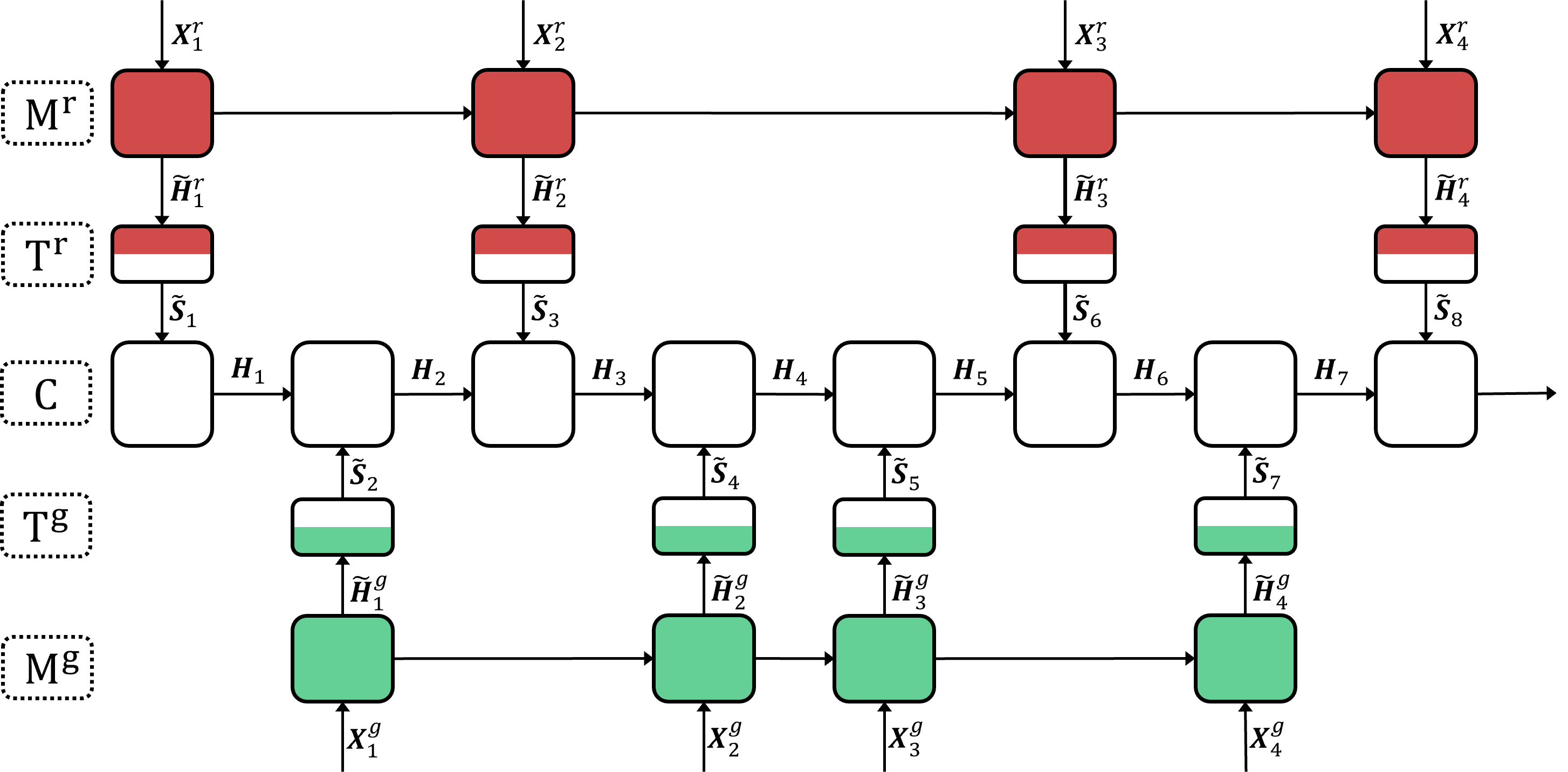}
\caption{Representation of the multiband ensemble of models. $\text{M}^{r}$ and $\text{M}^{g}$ represent single-band models. $\text{T}^{r}$ and $\text{T}^{g}$ correspond to the translation layers that project the data to a common representation. C is the central model with a similar architecture as shown in Figure \ref{fig:singleband_network}, trained on the outputs of the single-band models. The final classification is done with the output of the last step.}
\label{fig:MultiBand_network}%
\end{figure*}

We set the number of recurrent levels to two for {\em Gaia} and ZTF, and three for Pan-STARRS1, while maintaining the hidden state size consistent with that used in the single-band counterparts. The resulting multiband representation is captured in the vectors $\vec{H}$. Figure \ref{fig:MultiBand_network} provides an example diagram of the ensemble of models, depicting a two-band survey configuration.

\subsection{Color}\label{subsection:color}
In any astrophysical scenario, color information is critical. We understand color as the difference between two magnitudes, providing an approximation of an object's spectral shape. Typically, corrections due to extinction have to be applied to obtain the object's intrinsic color. Since the filters towards the blue part of the spectrum are most affected, objects appear redder than they truly are. In this work, we did not perform those corrections to make the pipelines simpler and not introduce any other source of uncertainty to the observations.

We transformed the single-band representations $\textbf{X}^b$ to obtain the mean magnitudes at each time step. From the magnitude differences, the magnitude at each time step is recovered. From there, the mean magnitudes in every band were obtained using forward fill imputation. The process is explained in Appendix \ref{appendix:Color}, which does not consider uncertainty propagation.  

\subsection{Final multiband representation}\label{subsection:final_representation}
The multiband time differences were included to provide the central model with a general view of the information. First, the single-band $\Delta {\rm MJD}$ were concatenated and sorted using the order information $\vec{o}^b$. Then, following the same method used for mean magnitudes detailed in Appendix \ref{appendix:Color}, the original MJD was recovered and used to compute the multiband time differences.

We appended the colors and time differences to the vectors $\vec{H}$ to provide more information about the object. The final representation was fed into a three-layer FFNN that featured batch normalization \citep{ioffe2015batch} and ReLU activation.

One of the main advantages of our multiband approach is its simplicity when making predictions. If a new observation arrives on any band the model was trained on—regardless of the number of bands—only two model evaluations are necessary: one for the observed band and another for the central model. 
The single-band models can be obtained by training band-specific models on multiple datasets or applying additional machine learning techniques, such as Transfer Learning.

\subsection{Training strategy}\label{subsection:training_strategy}
This study introduces a methodology involving $b+1$ interacting models, encompassing the single-band and a central unifying model. We applied backpropagation using the AdamW optimizer \citep{loshchilov2017decoupled}, coupled with an exponential decay learning rate of 0.95 every 60 training steps. We also employed early stopping with a patience of 20 epochs.

We independently trained each single-band network on the cross-entropy loss weighted by uncertainty, as depicted below:

\begin{equation}
\label{eq:loss_band}
L_{cls}(\vec{y}, \hat{\vec{y}}) = -\sum_{j=N_{skip}}^{N_i}{ w_j \sum_{k}{y_{k}\log{\hat{y}_{jk}}} },
\end{equation}
where $\hat{y}_{jk}$ are the predicted probabilities for the $k$-th class of the $j$-th step in the sequence and $y_{k}$ the ground truth, the same for the entire sequence. The class with the highest probability determined the final classification.

We trained the central network using the same classification loss. This network independently trained its weights, the weights associated with the $b$ linear combinations of hidden states as described in Eq. \ref{eq:elmo_sauce}, and the translation layers.

In this work, we aim to include the entire multiband observations in the model, as the single-band light curves are complementary. The information contained in the unified multiband representations is tied to the quality of the single-band embeddings. For machine learning models, the information contained on a light curve cannot be measured by using traditional features, as these models can extract information even from apparently noisy light curves \citep{2023A&A...670A..54D}. Additionally, the number of examples has a higher impact on the quality of the single-band representations than the noise characteristics \citep{Becker2020}.

As such, the central network is able to combine and enhance the provided information, even if the characteristics of the light curves are not ideal from a traditional standpoint. It is not in the scope of this work to explore the characteristics that a light curve might possess to be considered informative.

Moreover, comparing the quality of single-band light curves depends on many factors, such as the specific filters, extinction, different physical behaviors, or errors that might affect the hardware at different wavelengths. The criteria to balance the quantity and quality of the information will depend on the specific task to solve.

The uncertainties were used to weigh each observation's importance. Different bands can have drastically different uncertainty behavior. For simplicity, we ensured equal weighting of uncertainties across different bands to prevent undue prioritization of bands with smaller uncertainties. To achieve this, the uncertainties of each single-band light curve were normalized to add up to one. Then, we concatenated and sorted the values using the order information $\vec{o}^b$. Finally, we computed the loss weights using Equation \ref{eq:uncertainty}.

This strategy was intended to be used for surveys with a similar number of observations per band. In a different scenario, other strategies could be adopted, such as normalizing the single-band uncertainties to have a mean of one.

During each training batch, we performed backpropagation once in the single-band models and twice in the central model with half the learning rate.

The training process began with the single-band models, which we trained for 500 batches. We then start training the central model. This strategy allowed the single-band models to approximate a local minimum while still permitting minor parameter adjustments in the generation of their embeddings. These slight modifications in the input helped the central model alleviate overfitting. Allowing the single-band models to converge before training the central one resulted in poorer performance.

The number of trainable weights for the {\em Gaia} model was \num{383113} for the single-band models and \num{1198114} for the central representation, totaling \num{1964340}.
For Pan-STARRS1, the single-band models contained \num{110153} trainable parameters, and the central model \num{744645}, totaling \num{1295410}. For ZTF, these are \num{383500} for the single-band models, and \num{1197993} for the central model, totaling \num{1964993}.

\subsection{Multi-task training}\label{subsection:multitask_learning}
To obtain more general embeddings, we trained the central network for classification using cross-entropy and mean squared error (MSE) for regression of physical parameters.

For the regression task, only the last hidden state of the central model was used to predict the physical parameters. We created the final representation using a two-layer FFNN with batch normalization and ReLU activation.

The final loss function $L$ is the sum of the classification and regression losses for one example is
\begin{eqnarray}
    L_{reg}(\vec{y}, \hat{\vec{y}})  &=& \sum_{k}^{N_{p}}{\delta_{k}\lambda_k \cdot  {\rm MSE}(P_k, \hat{P}_k)},\\
    L &=& L_{cls} + L_{reg}\label{eq:loss_central}.
\end{eqnarray}
The regression loss, $L_{reg}$, was calculated for each physical parameter using the MSE, weighted by $\lambda_k$, which represents the relative importance of each parameter. 
Here, $N_p$ is the number of physical parameters, and $P_k$ is the prediction for the $k$-th physical parameter. Without proper weighting, the different dynamic ranges of the parameters could bias the learning process.

Since not all objects have multiband light curves or measurements for all physical parameters, we use the parameter $\delta_k$ to mask the missing ground truth values from the loss update. It took the value of 1 when the physical parameter had a valid measurement and 0 otherwise. This ensured that the model could be trained with all available information while accounting for the missing data.

The same final representation discussed in Sect. \ref{subsection:final_representation} was used for the regression branch. It was fed into a two-layer FFNN with 128 neurons per layer, employing ReLU activation and batch normalization after each layer. This architecture was applied consistently across all datasets and physical parameters.

\section{Baselines}\label{section:Baselines}
This section introduces the baselines used as points of comparison for the proposed model. The first baseline is the BRF, which is based on the work of \citet{sanchez2021alert}. This model is already utilized in the real-time classification of variable objects, making it a suitable candidate for comparison. A detailed description of the BRF model is provided in Section \ref{subsection:RandomForest}.

To assess the advantages of our proposed model's architecture and to evaluate an additional baseline, we utilized the concatenation of the representations generated by the single-band RNN models, as discussed in Sect. \ref{subsection:single-band}. This baseline approach allowed us to investigate the performance of a straightforward combination of single-band representations without the additional modeling and training steps introduced in our proposed model.

\subsection{RF baseline}\label{subsection:RandomForest}
We trained a BRF classifier for classification tasks and a RF for regression, utilizing the Python packages Imbalanced-learn \citep{JMLR:v18:16-365} and scikit-learn \citep{scikit-learn}, respectively.

To provide the input for the baseline model, we used the code provided by \citet{sanchez2021alert}, computing a total of 105 features. However, not all of these features were used since our data contained only time, band, magnitude, and uncertainty measurements. We independently computed the color features as the subtraction of two mean magnitudes. We followed the same methodology, imputing missing values with $-\num{999}$.

We discarded the mean magnitudes and computed the periods with single-band light curves to avoid biasing the model, given that multiband period estimation yields superior results \citep{Mondrik_2015}. 

We trained two versions of the BRF classification baselines: one limiting the maximum number of objects per class to replicate the training conditions of our model, and another without any restrictions, to study if that limitation hinders the performance of the model. In this way, the performance could be compared directly.

For each experiment, we performed a grid search in five stratified folds to determine the optimal hyper-parameters, utilizing the scikit-learn library. We optimized the macro F-score for classification and MSE for regression. Our search spanned between 100 and 4000 trees. Initially, we explored tree counts starting at 100 and increasing in increments of 100, up to 500. Beyond this, the numbers began at 750 and increased by 250 until we reached 4000.

\subsection{Mixture of experts}\label{section:mixtureofexperts}
We evaluated the merits of our proposed ensemble model in comparison to a similar architecture without the central network. Each single-band LSTM in this setup was constructed as detailed in Section \ref{subsection:single-band}.

The resultant embeddings from these single-band LSTMs were concatenated to form a multiband representation. Subsequently, we appended the color information to this multiband representation. The training process used the same data and hyper-parameters as those used in our proposed model. 

All of the RNNs were trained together as a single network, since the multiband observations are not simultaneous. The gradient backpropagation was performed from a single classification loss into each of the single-band RNNs. 

\section{Results }\label{section:Results}

This section presents the results of the proposed multiband model and the baselines. We split the data into $70\%$ for training, $10\%$ for validation, and $20\%$ for testing. We perform seven of such stratified splits, maintaining the same test set. We use the same data for our model and the baselines. For the RF-based models, the training and validation splits are combined, as these models do not need validation data. 

\subsection{Baselines}\label{subsection:baselines}
A grid search was performed for both classification and regression tasks, and the optimal number of trees was selected. For the two BRF experiments, the number of trees for {\em Gaia} was 2750 and 2750; for Pan-STARRS1, 1000 and 200; and for ZTF, 3250 and 3500, for the capped and uncapped experiments, respectively.

Upon evaluation on the test set, the capped experiment achieved median macro-averaged recall values of \num{0.800}, \num{0.420}, and \num{0.778} for {\em Gaia}, Pan-STARRS1 and ZTF, respectively. The uncapped experiment attained median macro-averaged recall values of \num{0.801}, \num{0.417}, and \num{0.779}, respectively.

In the RNN baseline model, the median macro-averaged recall values were \num{0.520} for {\em Gaia}, \num{0.358} for Pan-STARRS1, and \num{0.516} for ZTF.

The F-score for each class is detailed in Tables \ref{tab:gaia_baseline_results}, \ref{tab:ps_baseline_results}, and \ref{tab:ztf_baseline_results}.

We used only the datasets with a capped maximum number of objects per class for the regression baseline. The aim is to evaluate how much information can the RF and MTL models extract from the data to predict the measured physical parameters. The number of trees was calculated separately for each physical parameter. 

For {\em Gaia}, 2250 and 400 trees were used for $T_{\rm eff}$ and radius, respectively. Pan-STARRS1 used 3000 and 200, while ZTF used 3250 and 2250 trees for these parameters. 
The results are in Tables \ref{tab:R2_results}, \ref{tab:RMSE_results} and \ref{tab:MAPE_results}. The best results are highlighted in bold font.

Post-training feature importance analysis, obtained as the normalized total reduction of the Gini impurity, highlighted the significant role of color information in the RF. 
For {\em Gaia}, color features accounted for $84\%$ and $63\%$ of the importance in the $T_{\rm eff}$ and radius predictions, respectively. 
For Pan-STARRS1, color features held a combined importance of $75\%$ for $T_{\rm eff}$ and $79\%$ for radius predictions.
For ZTF, color features held a combined importance of $62\%$ for $T_{\rm eff}$ and $56\%$ for radius predictions.

\subsection{Multiband RNN model}\label{subsection:RNN_classifier}

Our proposed multiband ensemble yielded median macro-averaged recall values of \num{0.745} for the {\em Gaia} data, \num{0.588} for Pan-STARRS1, and \num{0.828} for ZTF. The median F-scores for each class, obtained in the seven cross-validation folds, are presented in Tables \ref{tab:gaia_baseline_results}, \ref{tab:ps_baseline_results}, and \ref{tab:ztf_baseline_results}. In the tables, Base denotes the baseline RNN model, Multi corresponds to our multiband model, and MTL signifies the multi-task learning model. The terms BRF and $\text{BRF}_{\text{All}}$ pertain to the models trained on capped and uncapped training sets, respectively. We highlight in bold the highest results per class. Notably, across all categories and throughout all three surveys, our ensemble of RNNs outperforms the Base model. Moreover, it generally surpasses the BRF models, with only two exceptions - the T2CEP in {\em Gaia} and the SNIa class for ZTF.

\begin{table}
  \caption[]{F-score for {\em Gaia} experiments results.}
     \label{tab:gaia_baseline_results}
\centering                   
\begin{tabular}{lrrrrrr}      
\hline\hline                 
        \text{Class}& Support& Base& Multi & BRF& $\text{BRF}_{\text{All}}$ & MTL\\
\hline                                   
        RRab &8000 & 0.82&\textbf{0.89}&0.86&0.86&0.88\\
        RRc& 3451 & 0.59&\textbf{0.76}&0.72&0.72&0.75\\
        MIRA\_SR& 8000 &0.99&\textbf{1.00}&\textbf{1.00}&\textbf{1.00}&\textbf{1.00}\\
        DSCT\_SXPHE& 680 &0.08&0.70&0.70&0.70& \textbf{0.71}\\
        CEP& 1166 & 0.56&\textbf{0.71}&0.69&0.70&\textbf{0.71}\\
        T2CEP& 236 & 0.00& 0.53&0.65&\textbf{0.66}&0.51\\      
\hline                                             
\end{tabular}
\end{table}

\begin{table}
  \caption[]{F-score for Pan-STARRS1 experiments results.}
     \label{tab:ps_baseline_results}
\centering                   
\begin{tabular}{lrrrrrr}      
\hline\hline                 
        \text{Class}& Support& Base& Multi & BRF& $\text{BRF}_{\text{All}}$ & MTL\\
\hline                                   
        RRab &  2000  &0.56&\textbf{0.78}&0.60&0.61&\textbf{0.78}\\
        RRc &   2000  &0.60&0.75&0.03&0.11&\textbf{0.76}\\
        RRd &   53  &0.00   &\textbf{0.06}&0.00&0.02&\textbf{0.06}\\
        MIRA\_SR &  788  &0.89&\textbf{0.95}&0.80&0.72&\textbf{0.95}\\
        DSCT\_SXPHE& 381 &0.01&\textbf{0.70}&0.64&0.57&0.68\\
        T2CEP&   38 &0.00&\textbf{0.36}&0.00&0.00&0.28\\
\hline                                             
\end{tabular}
\end{table}

\begin{table}
  \caption[]{F-score for ZTF experiments results.}
     \label{tab:ztf_baseline_results}
\centering                   
\begin{tabular}{lrrrrrr}      
\hline\hline                 
        \text{Class}& Support& Base& Multi & BRF& $\text{BRF}_{\text{All}}$ & MTL\\
\hline                                   
        AGN     &484 &0.08  &\textbf{0.63}&0.51&0.52&0.60\\
        Blazar  & 218&0.00  &\textbf{0.64}&0.40&0.39&0.58\\
        CV/Nova &152 &0.05  &\textbf{0.81}&0.53&0.53&0.73\\
        E       & 2000&0.79  &\textbf{0.93}&0.79&0.79&0.91\\
        LPV     &1863 &0.94  &\textbf{1.00}&0.92&0.93&0.99\\
        QSO     & 2000&0.83  &\textbf{0.90}&0.82&0.82&0.89\\
        RRL     & 2000&0.80  &\textbf{0.94}&0.75&0.76&0.91\\
        SNIa    & 101&0.74  &0.97&\textbf{0.98}&0.97&0.96\\
        YSO     &159 &0.05  &\textbf{0.80}&0.46&0.46&0.75\\
\hline                                             
\end{tabular}
\end{table}

\begin{figure}
\centering
\hspace{-1cm}
\includegraphics[width=0.5\textwidth]{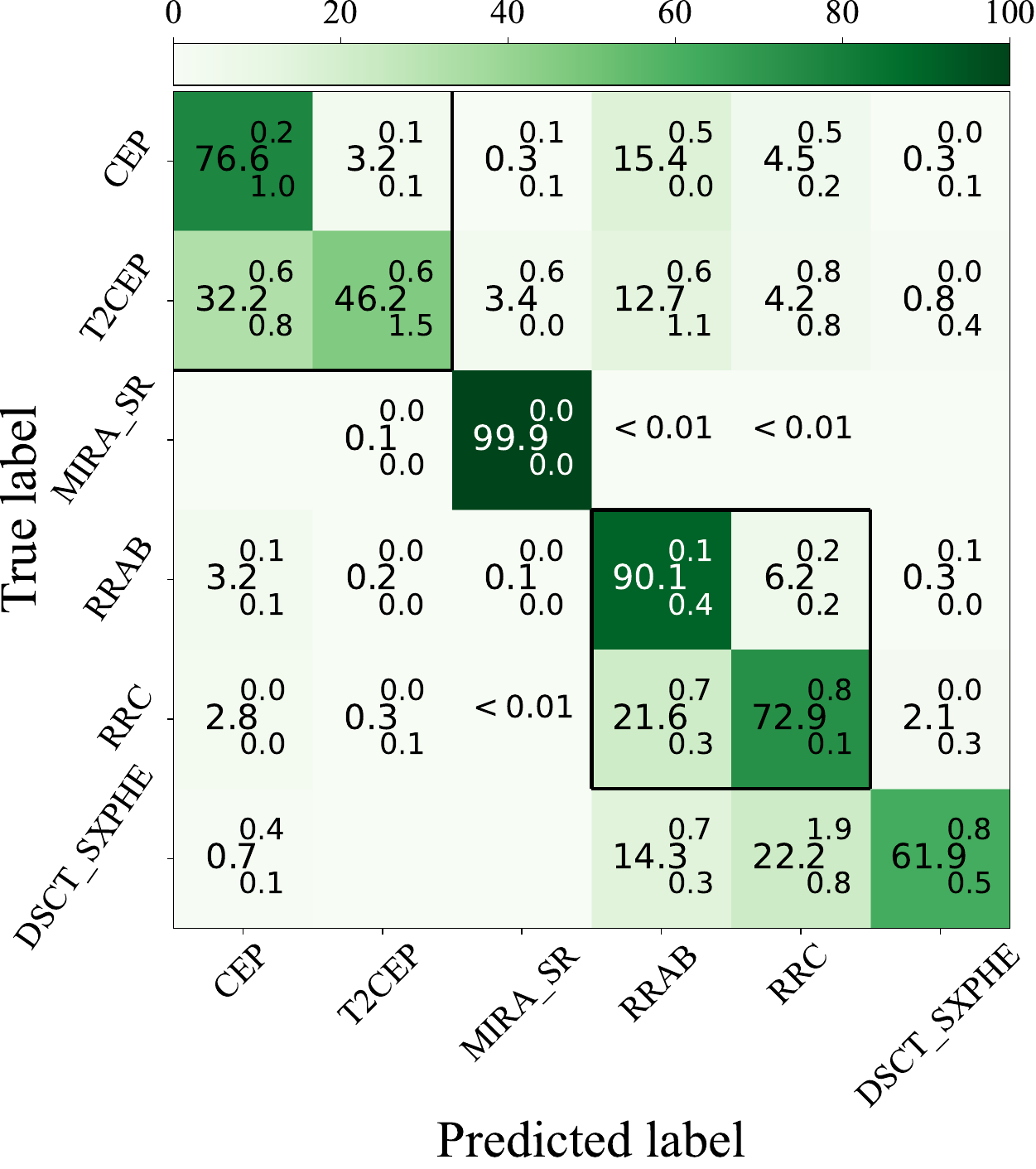}
\caption{Confusion matrix for the {\em Gaia} Multi model. Cepheids and RR Lyrae are confused inside the hierarchy. A bias towards the RRab class can be seen, as this class dominates the dataset. }
\label{fig:rnn_gaia}%
\end{figure}

\begin{figure}
\centering
\hspace{-1cm}
\includegraphics[width=0.5\textwidth]{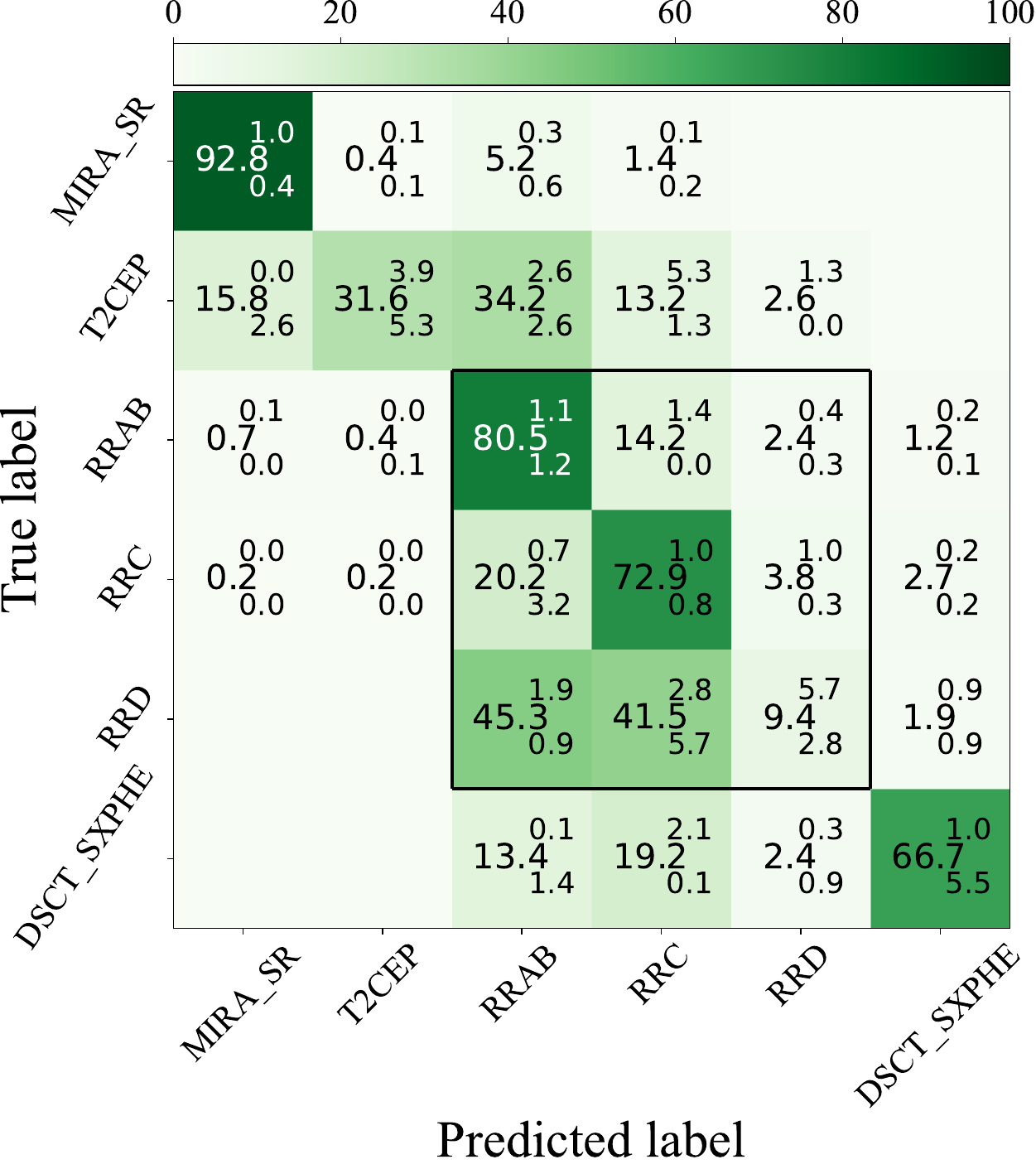}
\caption{Confusion matrix for the Pan-STARRS1 Multi model. RRd stars pulsate in the fundamental and first overtone simultaneously, with the latter being the dominant mode most of the time \citet[][and references therein]{2022MNRAS.517.5368B}. The same behavior can be seen in the fraction of misclassified RRd stars, which are assigned either RRab or RRc labels. }
          \label{fig:rnn_panstarrs}%
\end{figure}

\begin{figure}
\centering
\includegraphics[width=0.5\textwidth]{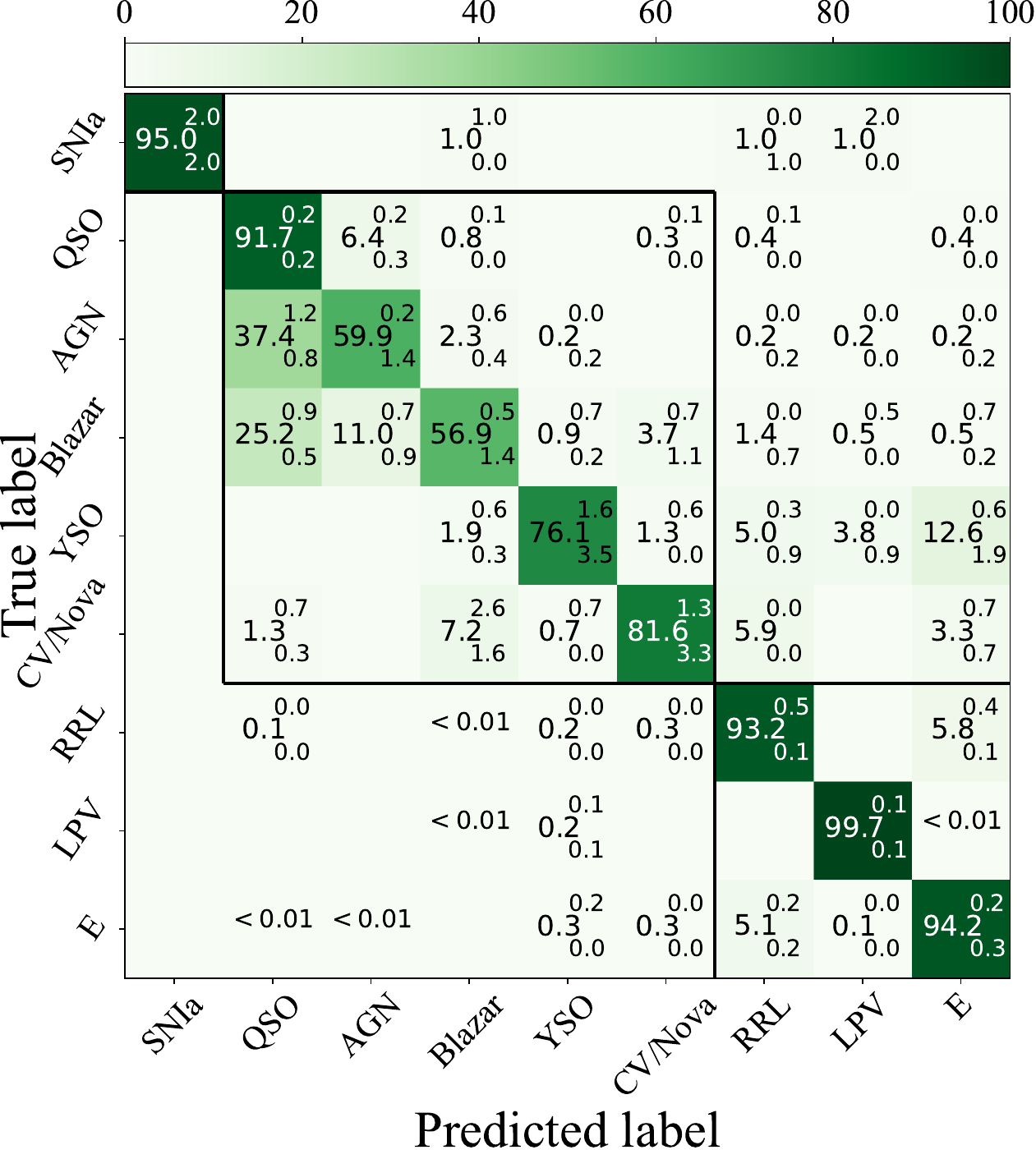}
\caption{Confusion matrix for ZTF Multi model. The main mistakes are between QSO, AGN and Blazars, given their similar source and a bias towards QSO as one of the most numerous classes. }
          \label{fig:rnn_alerce}%
\end{figure}

Figures \ref{fig:rnn_gaia}, \ref{fig:rnn_panstarrs}, and \ref{fig:rnn_alerce} display confusion matrices, showcasing the median results among the seven trained models. The rows do not sum to one, as the information is derived from different matrices. The values at the lower and upper end represent the 25th and 75th percentiles, respectively, with black lines grouping objects of the same type. The matrices are primarily diagonal, suggesting that most mistakes occur within larger classes.

As for the multi-task learning setup, the macro-averaged recall values for {\em Gaia} are \num{0.742}, \num{0.582} for Pan-STARRS1, and \num{0.807} for ZTF. The outcomes of the physical parameters prediction are depicted in Tables \ref{tab:R2_results}, \ref{tab:RMSE_results} and \ref{tab:MAPE_results} for the R2, root mean square error (RMSE) and mean absolute percentage error (MAPE) metrics, respectively.

For clarity purposes, we report the approximated value of $T_{\rm eff}$ to the nearest integer and the radius to the first decimal since the uncertainties generally fall within this order of magnitude. For training purposes, the data was used with all the reported decimals.

\begin{table}
  \caption[]{Regression R2 metric for the {\em Gaia}, Pan-STARRS1 and ZTF datasets.}
     \label{tab:R2_results}
\centering                   
\begin{tabular}{p{0.25\linewidth}lrr}     
\hline\hline                 
        Dataset &Model&  $T_{\rm eff}$ & Radius \\
\hline                                   
        \multirow{2}{*}{\text{{\em Gaia}}} &\text{MTL}& 0.860&0.411\\
                                    &\text{RF} & \textbf{0.909} & \textbf{0.785}\\
        &&&                                \\
        \multirow{2}{*}{\text{Pan-STARRS1}}  &\text{MTL}&0.760& \textbf{0.768} \\
                                    &\text{RF} &\textbf{0.793} &  0.774 \\
        &&&                                \\                                                                        
        \multirow{2}{*}{\text{ZTF}} &\text{MTL}& 0.739&\textbf{0.757}\\
                                    &\text{RF} & \textbf{0.753} & 0.661\\
            
\hline                                             
\end{tabular}
\end{table}

\begin{table}
  \caption[]{Regression RMSE metric for the {\em Gaia}, Pan-STARRS1 and ZTF datasets.}
     \label{tab:RMSE_results}
\centering                   
\begin{tabular}{p{0.25\linewidth}lrr}     
\hline\hline                 
        Dataset &Model&  $T_{\rm eff}$ & Radius \\
                &     &(K)      &($\text{R}_\odot$)\\        
\hline                                   
        \multirow{2}{*}{\text{{\em Gaia}}} &\text{MTL}& 467& 33.5\\
                                    &\text{RF} & \textbf{371} &\textbf{15.3} \\
        &&&                                \\
        \multirow{2}{*}{\text{Pan-STARRS1}}  &\text{MTL}& 530& 8.7\\
                                    &\text{RF} &\textbf{492} & \textbf{8.6}  \\
        &&&                                \\
        \multirow{2}{*}{\text{ZTF}}  &\text{MTL}& 500& \textbf{9.6}\\
                                &\text{RF} &\textbf{486} & 11.3  \\                                    
\hline                                             
\end{tabular}
\end{table}

\begin{table}
  \caption[]{MAPE for the {\em Gaia} and Pan-STARRS1 datasets regressions.}
     \label{tab:MAPE_results}
\centering                   
\begin{tabular}{p{0.25\linewidth}lrr}     
\hline\hline                 
        Dataset &Model&  $T_{\rm eff}$ & Radius \\
\hline                                   
        \multirow{2}{*}{\text{{\em Gaia}}} &\text{MTL}&6.00&53.63\\
                                    &\text{RF} & \textbf{4.56} &\textbf{49.74}\\
        &&&                                \\
        \multirow{2}{*}{\text{Pan-STARRS1}}  &\text{MTL}&6.89&107.05\\
                                    &\text{RF} &\textbf{6.42}&\textbf{84.84}\\
        &&&                                \\
        \multirow{2}{*}{\text{ZTF}}  &\text{MTL}&6.27&\textbf{78.41}\\
                                &\text{RF} &\textbf{5.93}&90.18\\                                    
\hline                                             
\end{tabular}
\end{table}

\begin{figure*}
\centering
\includegraphics[width=17cm]{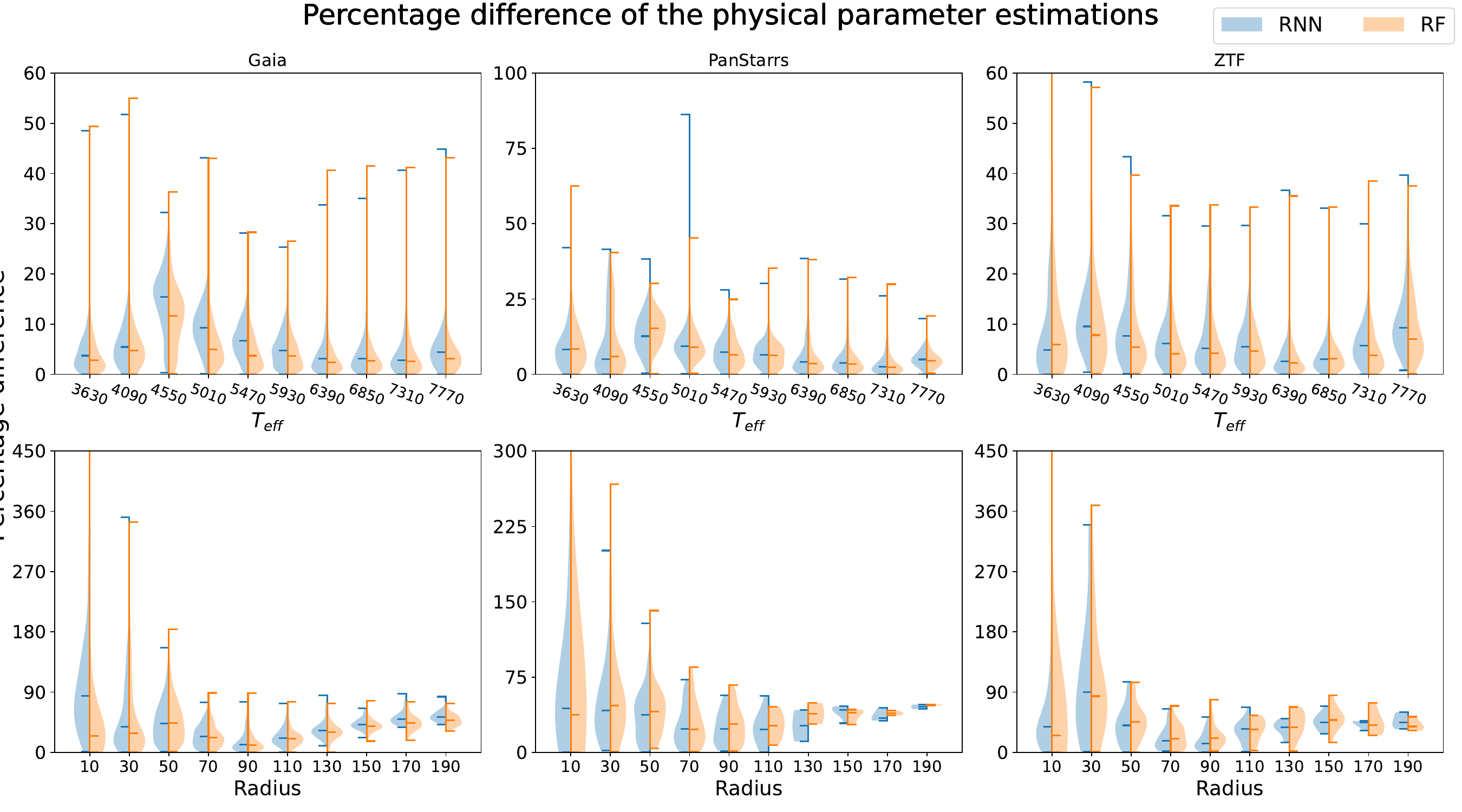}
\caption{Distribution of the physical parameter estimations. Each column corresponds to the parameters of a survey. }
          \label{fig:violin}%
\end{figure*}

To expand on the aggregated metrics, we report the distribution of predictions for our model and the RF in Fig. \ref{fig:violin}. It shows that the RNN systematically predicts similar but shifted distributions compared to the RF.

\subsection{Sequence classification}
The classification performance through time is important for any survey, as the classifier should provide stable and more reliable classifications as new observations arrive. Inspecting the model's predictions as a function of the observation number provides evidence of the overall performance. 

\begin{figure}
\centering
\includegraphics[width=0.47\textwidth]{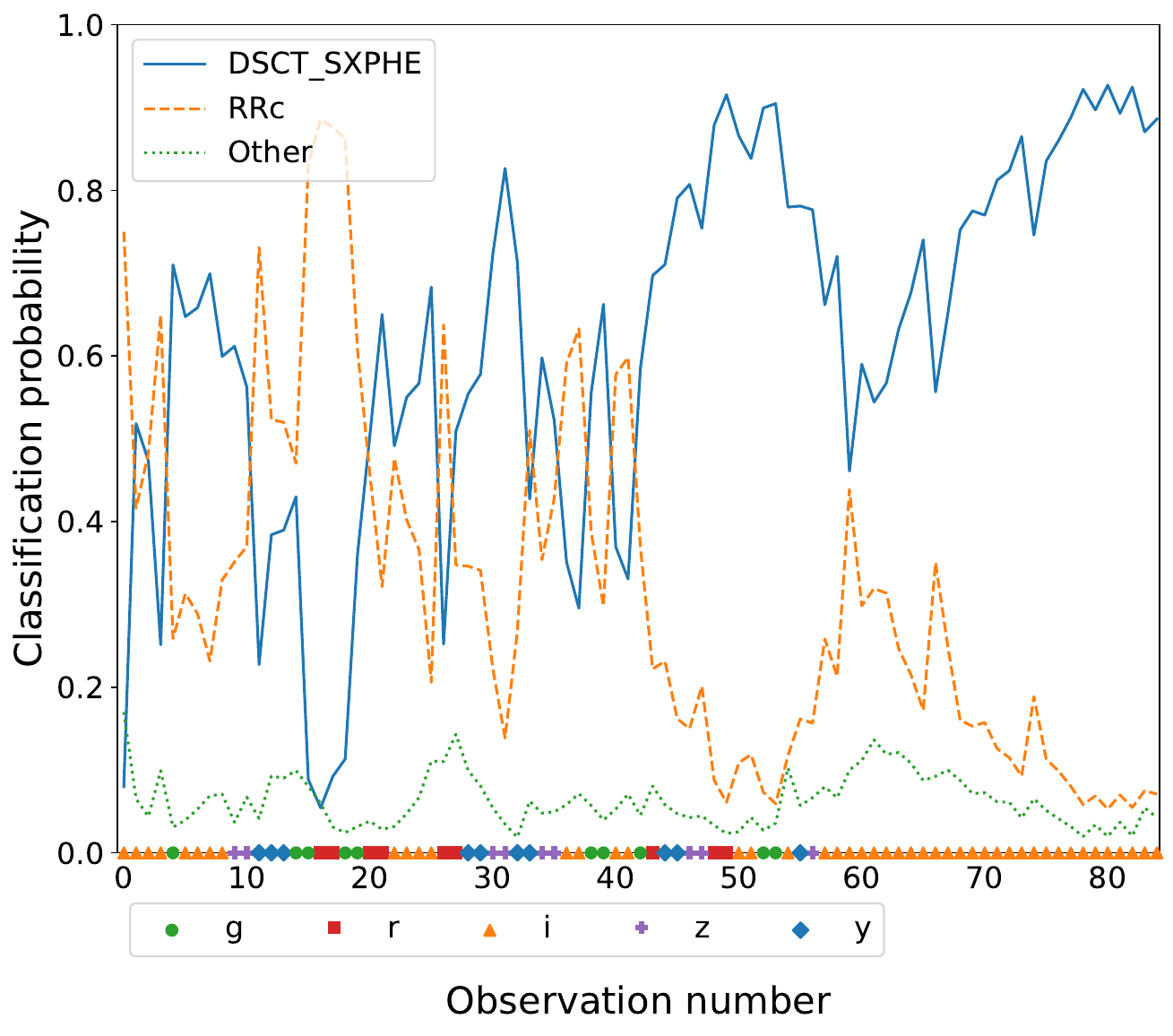}
\caption{Example of the models' output of the classification probabilities as a function of the number of observations for the Pan-STARRS1 DSCT\_SXPHE star 86363087019643005. The symbols on the lower axis identify the filters used for each observation. Filter \textit{g} is represented with a green circle, \textit{r} with a red square, \textit{i} with an orange triangle, \textit{z} with a purple cross, and \textit{y} with a blue diamond.
The model's main confusion was among DSCT\_SXPHE (blue solid line) and the RRc class (orange dashed line). The rest of the classes, represented as Other (green dotted line) are assigned a low probability.}
\label{fig:classification_time}%
\end{figure}

Figure \ref{fig:classification_time} presents the classification evolution with respect to the number of multiband observations for a single DSCT\_SXPHE object in the Pan-STARRS1 data. Initially, classification fluctuated between DSCT\_SXPHE and RRc classes, while the other classes maintained a low probability. After approximately 40 observations, the classification probability improves with new data, regardless of the band. At the final step, the predicted class stabilizes, but this behavior may vary for different light curves.

Our data representation precludes predictions for the first two observations in each band. Therefore, the plot demonstrates the performance based on our representation and does not precisely represent the actual performance as an alert classification mechanism.

\subsection{Classification performance}
\begin{figure}
\centering
\includegraphics[width=0.45\textwidth]{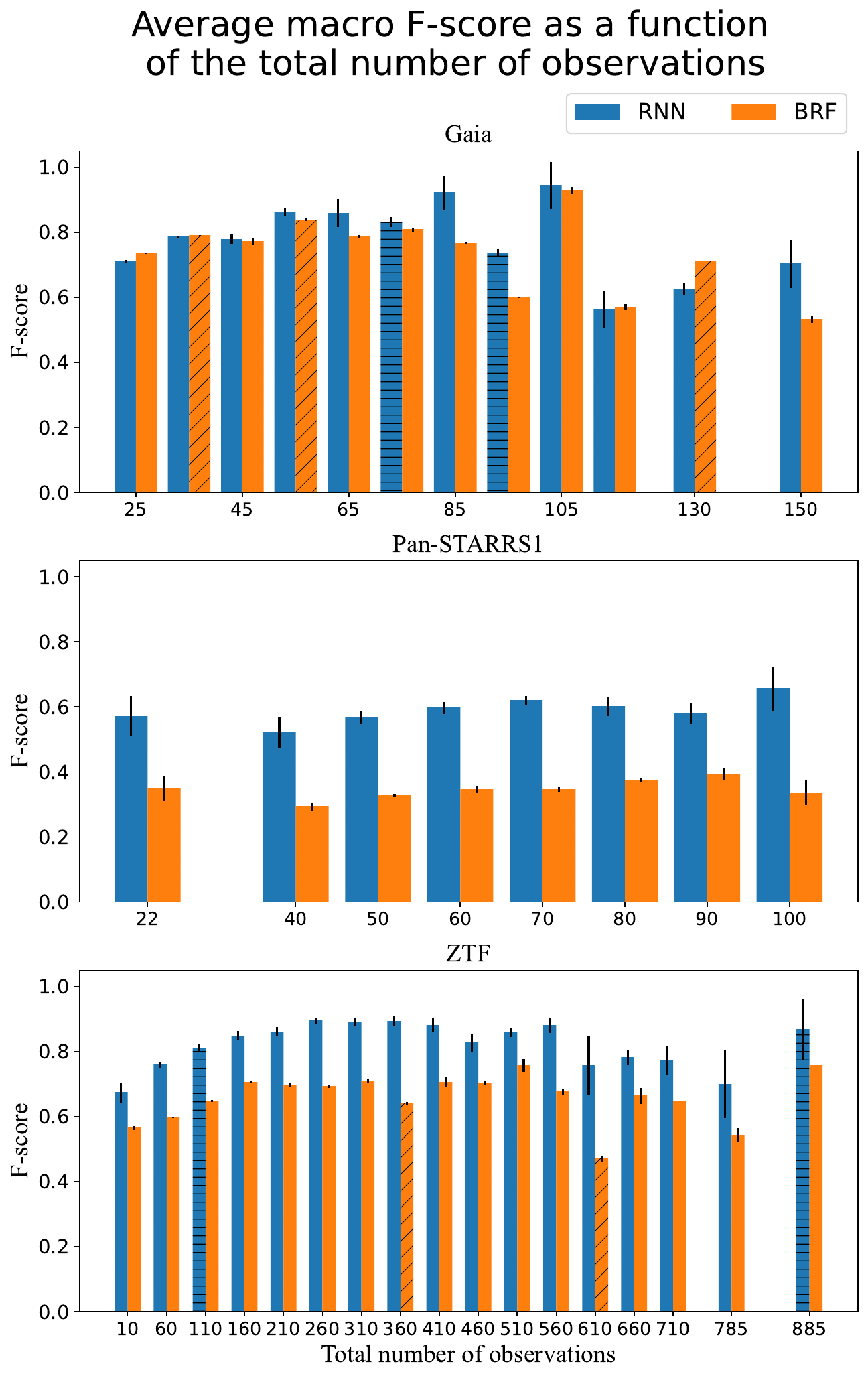}
\caption{F-score as a function of the total number of observations. The results of the RNN model are shown in solid blue, while the values for the BRF are shown in diagonal hatched orange. The bins are not uniform as some of them were merged to increase the number of objects per bin. Bins with less than 20 examples were removed to avoid showing unrepresentative results. The vertical black lines show the standard deviation for each bin.
}
\label{fig:cls_vs_N}%
\end{figure}

Feature-based classifiers depend on informative features, such as the period, which cannot be calculated when only a few observations exist. However, the sequential nature of an RNN-based classifier enables classification even with a limited number of observations per band. Figure \ref{fig:cls_vs_N} provides histograms of the mean F-score in bins based on the total number of observations.
Generally, the RNN model outperforms the BRF. The significant difference in Pan-STARRS1's performance can be attributed to the large number of bands, which shortens the single-band light curve lengths and consequently diminishes the quality of the features.

\section{Analysis}\label{section:Analysis}

Analysis of our multiband ensemble model reveals its adaptability across varied surveys, signified by its performance with different numbers of bands and cadences. This adaptability is anchored in the model's modular architecture, composed of single-band networks that function independently but are united by another central network. This independence allows the models to be individually trained on similar datasets, providing a considerable degree of flexibility.

The preprocessing used in this work is not refined for the early classification of transient objects, which requires a prediction from a small set of single-band observations. We note that the single-band models can be changed to a better-suited model, such as \citet{donoso2021effect}, which does not drop any observations.

The multiband RNN ensemble demonstrates a high degree of accuracy in multiband classification, as evidenced in Figures \ref{fig:rnn_gaia}, \ref{fig:rnn_panstarrs}, and \ref{fig:rnn_alerce} for {\em Gaia}, Pan-STARRS1, and ZTF data, respectively.

In the case of {\em Gaia}, a primarily diagonal confusion matrix indicates strong classification performance, with the model adeptly identifying main classes such as Cepheids and RR Lyrae. While the model shows a slight bias towards RRab objects, high-amplitude $\delta$ Scutis present a challenge, often being misclassified as RR Lyrae. In spite of the difference in their period distributions, the light curves of RRab objects may resemble that of a $\delta$ Scuti pulsating in the fundamental radial mode; in like vein, RRc light curves may resemble those of $\delta$ Scutis pulsating in the first radial overtone. This could point to the model's deficiency in characterizing objects with shorter periods, in contrast with its higher efficiency at longer periods, as revealed by the high scores that are found in the case of MIRA\_SRs.

The Pan-STARRS1 results mirror those from {\em Gaia}, with RRab stars being the most populous class and $\delta$ Scutis objects maintaining a similar proportion of misclassifications. A significant issue occurs with the RRd stars, most of which are misclassified due to the fact that they pulsate simultaneously in both the fundamental and first overtone, with the network assigning the class corresponding to the most dominant pulsation mode.

For the ZTF data, which extends to transients, stochastic, and periodic variables, the classification accuracy remains high. The primary confusion arises among QSO, AGN, and Blazar categories due to their shared source properties, also observed in \citet{sanchez2021alert}. 
Notably, a degree of overfitting is observed in Fig. \ref{fig:rnn_alerce}, as vertical patterns emerge from the predicted QSO and E classes. The model is biased towards those classes, as they are some of the largest categories in the dataset.
In specific cases, such as ZTF SNIa, the BRF outperforms our model, as some of the features are specifically designed to identify SNe.

In general, our multiband ensemble of models matches or surpasses the performance of the standard BRF, delivering higher F-scores across multiple surveys. Our proposed model outperforms the BRF in all experiments except {\em Gaia} T2CEP and ZTF SNIa, where the feature-based method achieves a better F-score. The differences are minimal, as the former is a class with few examples where both models perform poorly, and the latter shows a difference of 0.01 in F-score, which can be increased with careful hyper-parameter tuning.
The results underline the model's effectiveness and flexibility, making it a promising tool for future astronomical surveys. In terms of design, our model can adapt to any survey's cadence, regardless of the number of filters. This flexibility is illustrated in Fig. \ref{fig:classification_time}.

Our multiband model is specifically adept at providing solid classification performance even for light curves with few observations, as shown in Figure \ref{fig:cls_vs_N}. The BRF faced difficulties as not all features could be accurately computed, as the imputation scheme negatively impacts the classifier's performance as not all features are well-defined.

In the regime of longer light curves, our model continues to provide correct classifications, even for long sequences where the features are well defined, outperforming the baselines. 
However, as seen in Figure \ref{fig:cls_vs_N}, our model's performance does decrease slightly for longer sequences. This is a characteristic limitation of RNNs that tend to lose track of information over extended sequences \citep{pascanu2013difficulty}. However, for the expected number of observations from the Rubin/LSST, the results are stable, and no significant impact should be expected.

In the context of multi-task learning, the performance of the MTL model remains comparable to the pure classifier (Multi) model. In most instances, it manages to achieve similar performance, and in some classes even surpasses it, such as in the case of {\em Gaia} DSCT\_SXPHE and Pan-STARRS1 RRc. 
In some classes, the classification performance of the MTL model decreases, such as T2CEP stars in Pan-STARRS1, {\em Gaia} RRab class or the entire ZTF data. For the first two datasets, the differences are at most 0.02 in F-score (with the exception of Pan-STARRS1 T2CEP), and with proper hyper-parameter tuning, the score should at least equalize. For ZTF, the scores are consistently lower, but with much better physical parameter estimation.

We note that training the models for classification and regression simultaneously has the added advantage of improving the quality of the embeddings. These enriched embeddings can subsequently be leveraged to resolve other downstream tasks, such as the regression of other astrophysical quantities.

When it comes to the regression of physical parameters, the results in Tables \ref{tab:R2_results}, \ref{tab:RMSE_results}, and \ref{tab:MAPE_results} show the RF as the best model. Nonetheless, our model was able to extract representations used to perform classification and regression simultaneously. While RFs can be trained to perform individual tasks, they are built from a predefined representation of the objects, cannot extract a new one from the data, and do not benefit from the information of other tasks.
This highlights the importance of flexible models and informative embeddings that can be used as a foundation for more task-specific models \citep{touvron2023llama, 2023A&A...670A..54D}.

A plausible explanation for this could be the strong dependence of the parameters on the color information. Although both models had equal access to this information, the more straightforward approach adopted by RF appears to be better suited for this task. It is conceivable that a more specialized model could potentially match the RF results. The development of such a model extends beyond the purview of this study.

\section{Conclusion}\label{section:Conclusion}
The aim of this work was to develop an ensemble of RNNs capable of extracting informative representations from multiband light curves, in order to perform tasks such as classification of variable objects or regression of physical parameters.

The presented ensemble of multiband models was tested on real light curves from {\em Gaia}, Pan-STARRS1, and ZTF surveys, demonstrating its ability to adapt to varying numbers of bands. This unique model ingests multiple single-band light curves per object, eliminating the need for interpolation or binning, and lets the model itself extract a multiband representation.

Our model architecture allows it to maximize single-band information extraction, which is then combined into a unified multiband embedding, independent of each band's cadence. This gives our approach significant flexibility, as new single-band models can be trained on different datasets and included in the ensemble without requiring retraining of existing individual band models. Only the central RNN needs to be retrained to incorporate the new information.

Our ensemble approach demonstrates superior performance in scenarios with fewer observations, thus providing potential for early classification of sources from facilities such as Vera C. Rubin Observatory's LSST. Remarkably, even in cases like Pan-STARRS1, where six interacting models are trained, only two model evaluations are needed at prediction time: one for the observed band and one for the central representation. 

This method outperforms standard approaches, particularly those based on BRF implementations, especially in scenarios with more bands and sparser single-band light curves.

In addition, our research showed that a multi-task learning approach could enrich the embeddings obtained by the model. These improved embeddings can further be utilized in various astrophysical tasks, such as the regression of physical parameters like the temperature or radius of stars. We note that the features correlate strongly with the regressed parameters, which explains the edge of the RF. In this work, the multi-task learning approach increased the information contained in the embeddings. The regression task served as a regularization tool and was not fine-tuned for performance, as doing so would hurt the performance of the classifier, as seen in the ZTF case. Alternatively, it would require implementing a metric to balance the regression and classification tasks, which falls outside the scope of this work.

In future work, our model can be trained on more extensive datasets such as {\em Gaia} DR3, which offers improved photometry and longer light curves. In addition to this, we can incorporate metadata from an expanded array of sources, including but not limited to extinction maps, coordinates, and redshifts. This enriched data pool could significantly boost the model's predictive power and versatility. However, incorporating diverse datasets will require careful treatment to address potential systematics, inconsistencies, and errors inherent in the data, ensuring robust and reliable model performance.

Moreover, including a wider variety of tasks in our model—be it classification or regression—may further improve the quality of the embeddings. Such advancements could prove particularly beneficial in applications like determining the orbital parameters of binary systems or estimating parameters for diverse object types that extend beyond periodic ones. 

There is considerable potential to adopt advanced training techniques from the Natural Language Processing domain, such as self-supervised pre-training followed by supervised fine-tuning as in \citet{devlin2018bert}. This approach could enrich the learned data representations, paving the way for effective transfer learning, particularly considering the minor discrepancies in the filter systems used by ZTF, Pan-STARRS1, and Rubin/LSST.

Such strategies could facilitate model training on datasets from different surveys, such as ZTF, necessitating fewer training epochs for adaptation to Rubin/LSST conditions. This would considerably reduce the time required to develop a specific Rubin/LSST classifier, thus promoting efficiency and expediency in future astronomical studies.

\section{Data availability}
The code to create the data representation, and to train and test the multiband model is available in GitHub at \url{https://github.com/iebecker/ScalableMultiband_RNN}.

\begin{acknowledgements}
This research was supported by CONICYT-PFCHA/Doctorado nacional/2018-21181990. Additional support for this project is provided by ANID’s Millennium Science Initiative through grant ICN12\_009, awarded to the Millennium Institute of Astrophysics (MAS); by ANID/FONDECYT Regular grant 1231637; and by ANID's Basal grant FB210003.
\end{acknowledgements}
\bibliography{one}

\begin{appendix}

\section{Pan-STARRS1 cleaning}\label{appendix:PS_cleaning}
To clean invalid or noisy observations for the Pan-STARRS1 dataset, the following criteria is used, taken from Table 2 in \citet{magnier2013pan}:
\begin{itemize}
    \item \tt{psfQfPerfect}>0.9
    \item \tt{psfFlux>}0 and \tt{psfFluxErr}>0
    \item \tt{psfFlux}>\tt{psfFluxErr}
    \item \tt{infoFlag} $\not\in$ \{8, 16, 32, 128, 256, 1024, 2048, 4096, 8192, 32768, 65536, 131072, 262144, 4194304, 268435456, 536870912, 1073741824, 2147483648\}
    \item \tt{infoFlag2} $\not\in$ \{8, 16, 32, 64, 4096, 8192, 16384, 4194304\}
    \item \tt{infoFlag3} $\not\in$ \{8192, 16384\}
\end{itemize}

\section{Color information}\label{appendix:Color}
The process to obtain the color information from multiple single-band representations is described below. This process is computed at each time step and will be referred to as the cumulative mean magnitude. For simplicity, the process for one object is shown.

The magnitude differences are extracted from the last column of the matrix $\textbf{X}^b$ and from the third element of the first row. We form a vector whose components range from $\Delta m_2$ to $\Delta m_N$.
This vector is multiplied by a lower triangular matrix of ones, obtaining an expression containing the vector of magnitudes, shown below:
\begin{align}
\label{eq:mean_magnitude}
\begin{bmatrix}
1  &0 & 0& \cdots & 0\\
1  &1 & 0& \cdots & 0\\
1  &1 & 1& \cdots & 0\\
\vdots  &\vdots & \vdots&  & \vdots\\
1  &1 & 1& \cdots & 1\\
\end{bmatrix} 
\begin{bmatrix}
\Delta m_2\\
\Delta m_3\\
\Delta m_4\\
\vdots  \\
\Delta m_N\\
\end{bmatrix}
=
\begin{bmatrix}
m_2 - m_1\\
m_3 - m_1\\
m_4 - m_1\\
\vdots  \\
m_N - m_1\\
\end{bmatrix}
=
\begin{bmatrix}
m_2 \\
m_3 \\
m_4 \\
\vdots  \\
m_N \\
\end{bmatrix}
-
m_1\begin{bmatrix}
1 \\
1 \\
1 \\
\vdots  \\
1 \\
\end{bmatrix}.
\end{align}   

The vector of magnitudes can be obtained by adding to the latter the first magnitude $m_1$. This resulting vector is multiplied by another lower triangular matrix, which results in the sum of the magnitudes, as follows:
\begin{align}
    \label{eq:mean_magnitude_2}
    \begin{bmatrix}
    1  &0 & 0& \cdots & 0\\
    1  &1 & 0& \cdots & 0\\
    1  &1 & 1& \cdots & 0\\
    \vdots  &\vdots & \vdots&  & \vdots\\
    1  &1 & 1& \cdots & 1\\
    \end{bmatrix} 
    \begin{bmatrix}
    m_2\\
    m_3\\
    m_4\\
    \vdots  \\
    m_N\\
    \end{bmatrix}
    +
    \begin{bmatrix}
    m_1\\
    m_1\\
    m_1\\
    \vdots  \\
    m_1\\
    \end{bmatrix}
    =
    \begin{bmatrix}
    m_1 + m_2 \\
    m_1 + m_2 + m_3\\
    m_1 + m_2 + m_3+ m_4\\
    \vdots  \\
    \sum_{\bf{i=}1}^{N}{m_i}\\
    \end{bmatrix}.
  \end{align}  

To obtain the cumulative mean magnitudes, the result of Equation \ref{eq:mean_magnitude_2} is divided by the number of observations considered, 
\begin{align}
    \label{eq:mean_magnitude_3}
    \begin{bmatrix}
    \bar{m}_2\\
    \bar{m}_3\\
    \vdots \\
    \bar{m}_N\\
    \end{bmatrix} 
    =
    \begin{bmatrix}
    m_1 + m_2 \\
    m_1 + m_2 + m_3\\
    \vdots  \\
    \sum_{\bf{i=}1}^{N}{m_i}\\
    \end{bmatrix}
    /
    \begin{bmatrix}
    2 \\
    3\\
    \vdots  \\
    N\\
    \end{bmatrix}.
  \end{align}  
These operations can be computed efficiently using tensor operations on an entire batch of light curves. 

The observations are taken only in one band at a time. As our model necessitates the computation of all colors at every step in the sequence, we have to use an imputation method to compensate for missing mean magnitudes.

The imputation method involves propagating the most recently computed value of the mean magnitude. This propagated value is used until a new observation provides an updated mean magnitude. This method ensures continuity and offers a plausible fill-in for missing values, thereby maintaining the integrity of the model's performance.

This process can be visualized as follows, for a three-band example:
\begin{equation}
\label{eq:mean_colors}
\begin{bmatrix}
^1\bar{m}_2 & ^1\bar{m}_2 & ^1\bar{m}_2& ^1\bar{m}_3& ^1\bar{m}_3& ^1\bar{m}_3\\

            & ^2\bar{m}_2 & ^2\bar{m}_2& ^2\bar{m}_2& ^2\bar{m}_3& ^2\bar{m}_3\\

            &             & ^3\bar{m}_2& ^3\bar{m}_2& ^3\bar{m}_2& ^3\bar{m}_3\\
\end{bmatrix}.      
\end{equation}
\noindent Each row corresponds to single-band mean magnitudes, labeled from 1 to 3 superscripts. The subscript indicates the observation number on each band.
In this example, as the observations are taken sequentially, band 1 is updated first, then band 2, and finally band 3. The cumulative mean magnitudes are carried forward until a new observation is made available. In this case, the fourth observation is used to update the band 1 mean magnitude. 
\end{appendix}

\end{document}